\documentclass[11pt]{article} 
 
\usepackage{epsfig}  
\usepackage{amsmath}
\usepackage{amssymb}
\usepackage[all]{xy}

\newcommand{\D}{\displaystyle}

\DeclareMathOperator{\Iso}{Iso}
\DeclareMathOperator{\SO}{SO}

\DeclareMathOperator{\sech}{sech}
\DeclareMathOperator{\Rea}{Re}

\DeclareMathOperator{\supp}{supp}
\DeclareMathOperator{\id}{id}

\usepackage{amsthm}
\theoremstyle{definition}
\newtheorem{theorem}{Theorem}[section]
\newtheorem{lemma}[theorem]{Lemma}
\newtheorem{corollary}[theorem]{Corollary}
\newtheorem{proposition}[theorem]{Proposition}

\newtheorem{remark}[theorem]{Remark}

\begin{document} 

\vspace{2cm}

\title
{\vskip -70pt
\vskip 2cm
{\huge {Dynamics of $\mathbb{CP}^{1}$ lumps on a cylinder}}\\  
\vspace{8mm}
}

\author{
{\Large \bf Nuno M. Rom\~ao}
\thanks{e-mail: {\tt nromao@mpim-bonn.mpg.de}}\\[6mm]
{\normalsize \sl Max-Planck-Institut f\"ur Mathematik in 
den Naturwissenschaften}\\
{\normalsize \sl Inselstra{\ss}e 22, D--04103 Leipzig, Germany} \\[3mm]
{\normalsize \sl Max-Planck-Institut f\"ur Mathematik}\\
{\normalsize \sl Vivatsgasse 7, D--53111 Bonn, Germany}
\\}

\date{March 2004}

\maketitle

\begin{abstract}

\noindent
The slow dynamics of topological solitons in the $\mathbb{CP}^{1}$ 
$\sigma$-model, known as lumps, can be approximated by the geodesic 
flow of the $L^{2}$ metric on certain moduli spaces of holomorphic maps. 
In the present work, we consider the dynamics of lumps on an 
infinite flat cylinder, and we show that in this case the approximation can 
be formulated naturally in terms of regular K\"ahler metrics. 
We prove that these metrics are incomplete exactly in the multilump
(interacting) case. The metric for two-lumps can be computed in closed 
form on certain totally geodesic submanifolds using elliptic 
integrals; particular geodesics are determined and discussed in terms of 
the dynamics of interacting lumps.

\end{abstract}

%======================================================================

\section{Introduction}

Many field theories possess topological solitons as classical solutions,
and the study of their dynamics has long been an important research topic
in mathematical physics. Exact results for this problem
have only been obtained for rather special (integrable) models in $1+1$ 
dimensions; more generally, one has to resort to approximations
based on truncations of the field theories to finite-dimensional 
configuration spaces of collective coordinates. One such scheme is the
adiabatic approximation, first proposed by Manton in the context of BPS 
monopoles \cite{ManRk}. It has been
applied to extract detailed information about the slow dynamics of solitons 
in a number of models in $1+2$, $1+3$ and more dimensions (notably gauged 
Ginzburg--Landau vortices \cite{SamVS, StuAH} and Yang--Mills--Higgs 
monopoles \cite{AtiHitMM, StuYMH}) and is believed to work well for a large 
class of field theories exhibiting self-duality.

One neat example of a self-dual field theory is the nonlinear $\sigma$-model
with $\mathbb{CP}^{1}$ target on a Riemann surface $\Sigma$.
This is the dynamical system for maps
\[
W: \Sigma \rightarrow \mathbb{CP}^{1}
\]
described by the wave equation associated to a specific riemannian metric 
on $\Sigma$ and the usual round metric on the two-sphere $\mathbb{CP}^{1}$.
Topological solitons in this model, usually referred to as lumps, will 
typically arise if $\Sigma$ is compact or effectively compactified by 
suitable boundary conditions.
Static solutions are harmonic maps, with energy given by the usual Dirichlet 
integral. In the adiabatic approximation, one constructs another dynamical
system whose configuration space consists of the static solutions of minimal 
energy, the dynamics being defined by restricting the action functional of 
the original field theory. This space is stratified by homotopy classes,
and the strata are usually referred to as the moduli spaces.
For the $\mathbb{CP}^{1}$ $\sigma$-model, 
the Dirichlet energy is minimized exactly by the holomorphic or 
antiholomorphic maps within each 
homotopy class, labelled by the Brouwer degree $n\in \mathbb{Z}$ of $W$.
The moduli spaces (if non-empty) then have the structure of 
finite-dimensional complex varieties \cite{Nam}, and the adiabatic dynamics 
is geodesic motion with respect to a metric on them.
The Cauchy--Riemann equation, a first-order PDE, replaces the
second-order static equations of motion as a description of the fields.
This is the essential common feature to all self-dual theories. In the
adiabatic programme, the moduli spaces are often smooth manifolds equipped 
with natural geometric structures (symplectic forms, metrics of special 
holonomy) that turn out to be interesting objects by themselves. In some 
instances, they have even been used to probe aspects of the quantum field 
theories underlying the original models \cite{GibManCQM, SchQSM, RomQCS}.

The $\mathbb{CP}^{1}$ $\sigma$-model has applications to the physics of
ferromagnets and as a high-energy effective model for vortices; however,
its main interest has been as a toy-model displaying many of the 
features of more important field theories with gauge symmetry.
The adiabatic approach to this model was first investigated by 
Ward for the case
$\Sigma=\mathbb{R}^{2}$ in \cite{WarSML}; he found
that the approximation
is ill-defined, in the sense that the metric is infinite along certain 
directions that appear as frozen degrees of freedom. 
One way to regularise the metric is to place the vortices on a compact 
surface, and this was studied by Speight
when $\Sigma$ is a sphere \cite{SpeS2} or a torus  \cite{SpeT2}. It has also
been found that the metric for $\Sigma=\mathbb{R}^{2}$ regularises once
a self-gravitating interaction is included in the lagrangian \cite{SpeStr}.
Determining these metrics in closed form is in general beyond reach, but
some explicit formulae have been obtained in a number of nontrivial cases, 
namely 
for one-lumps on $\Sigma=S^2$ \cite{SpeS2} and for certain totally geodesic 
submanifolds of two-lumps on $\mathbb{R}^{2}$ \cite{WarSML} and on the 
particular torus  $\mathbb{C}/(\mathbb{Z}\oplus i \mathbb{Z})$ \cite{SpeT2}. 
Geodesic incompleteness of the moduli spaces was proved in \cite{SadSpe}.
There is also a general belief that the relevant metrics should be K\"ahler 
\cite{DinZakK, RubSusy}; this has been rigorised for $\Sigma=S^{2}$, 
and for $\Sigma=T^{2}$ and $n=2$ \cite{SpeL2}. The accuracy of the adiabatic 
approximation has been studied recently by Haskins and Speight 
\cite{HasSpe} in the spirit of work by Stuart \cite{StuAH, StuYMH} on 
the gauge theory models.

In this paper, the adiabatic dynamics of $\mathbb{CP}^{1}$ lumps is studied in 
some detail for the case where $\Sigma$ is an infinite cylinder. 
We can say that
this is an intermediate case between the situations $\Sigma=\mathbb{C}$
and $\Sigma$ compact considered by previous authors.
In the former, the metrics are ill-defined but explicit calculations of
the metric are possible, whereas in the latter the metrics are regular
but extremely hard to compute; the cylinder turns out to combine the
advantages of both. So our study complements
the existing literature in a setting that is unifying in some way, and our
results will reflect this. Let us summarise how this paper is organised.
We use the next section to fix the basic notation.
In section~\ref{secmsl}, we obtain elementary properties of the moduli 
spaces; we formulate the adiabatic approximation in terms of regular
riemannian metrics, which are shown to be K\"ahler. 
In section~\ref{secisom}, we discuss the isometries of these metrics.
The one-lump sector is studied in section~\ref{secc1l}. We then
establish that all the multilump metrics are incomplete in
section~\ref{secincomp}.
In section~\ref{secdc2l}, we address the two-lump dynamics and derive
more explicit results about the metric, its geodesics and curvature
properties. Finally, we discuss our results in section~\ref{secdisc}.

%=======================================================================

\section{The $\mathbb{CP}^{1}$ $\sigma$-model on a cylinder}
\label{secCP1oncyl}

For the rest of the paper, we shall take $\Sigma$ to be the infinite cylinder
\[
\Sigma=\mathbb{C} /(2 \pi i \mathbb{Z})
\]
with local complex coordinates $z=x+iy$ and metric
\begin{equation} \label{metricSig}
ds^{2}_{\Sigma} = dx^2 + dy^{2}
\end{equation}
induced from the euclidean metric of its universal cover $\mathbb{C}$.

The action for the $\mathbb{CP}^{1}$ $\sigma$-model, whose objects are
differentiable maps $W: \Sigma \rightarrow \mathbb{CP}^{1}$ dependent on 
time $t$, is 
given by
\begin{equation} \label{action}
I[W]=\int_{\mathbb{R}}(T-V)\, dt
\end{equation}
with kinetic and potential energies
\begin{eqnarray}
T & := & {2} \int_{\Sigma} \frac{|\dot{W}|^{2}}
        {\left( 1+ |W|^{2} \right)^{2}} d\mu_{\Sigma},  \label{Tenergy} \\
V & := & {4} \int_{\Sigma} \frac{|\partial_{z}W|^{2}+
                             |\partial_{\bar{z}}W|^{2}}
        {\left( 1+ |W|^{2} \right)^{2}} d\mu_{\Sigma}.  \label{Venergy}
\end{eqnarray}
We shall only consider the dynamics of maps $W$ for which the potential
energy $V$ above is finite. We represent $W$ by means of an inhomogeneous 
coordinate taking
values in $\mathbb{C} \cup \{ \infty \}$ following usual practice; overdots 
denote time derivatives and $d\mu_{\Sigma}$ is
the measure on $\Sigma$ associated to (\ref{metricSig}). The 
variational principle yields the wave equation as equation of motion, 
and static solutions are harmonic
maps from $(\Sigma,ds^{2}_{\Sigma})$ to $(\mathbb{CP}^{1},ds^{2}_{S^{2}})$.
Here, $ds^{2}_{S^{2}}$ is the riemannian metric on $\mathbb{CP}^{1}$
regarded as a two-sphere of unit radius; the K\"ahler $(1,1)$-form of
this metric will be denoted by $\omega_{S^{2}}$.
Following an argument first presented by Belavin and Polyakov \cite{BelPol}
(but already observed in a more general setting in the
mathematical literature --- cf.~\cite{Lic} p.~374), we write for a map 
$W:\Sigma \rightarrow \mathbb{CP}^{1}$
\begin{eqnarray*} 
0 & \le & {2} 
\int_{\Sigma} \frac{|(\partial_{x}\pm i\,\partial_{y}) W|^{2}}
{\left( 1+ |W|^{2} \right)^{2}} d\mu_{\Sigma} \\
 & = & 4 \int_{\Sigma} \frac{|\partial_{z}W|^{2}+|\partial_{\bar{z}}W|^{2}}
        {\left( 1+ |W|^{2} \right)^{2}} d\mu_{\Sigma} \mp
        \int_{\Sigma} W^{*}(\omega_{S^{2}}) \\
 & = & V[W] \mp  {\rm deg}(W) \, {\rm Vol}(S^{2}),
\end{eqnarray*}
where $\deg (W)$ is the Brouwer degree of the map. In the inequality above,
it is useful to take the top signs if $n:=\deg (W)$ is nonnegative, and
the bottom signs otherwise. Then we learn that
\[
V[W] \ge  4 \pi |n|,
\]
which implies that $n\in \mathbb{Z}$ provided $V$ is finite. Moreover, we 
deduce that the potential (or Dirichlet) energy (\ref{Venergy}) is
minimized to $4 \pi |n|$ on each topological class by a solution of 
the Cauchy--Riemann equation
\begin{equation} \label{CReq}
\partial_{\bar{z}} W =0
\end{equation}
if $n \ge 0$, or $\partial_{z}W=0$ if 
$n<0$. To simplify our discussion, we will mostly be considering the
case $n \ge 0$ only, but all the statements can be easily adapted to
the $n < 0$ case.

In this paper, we shall be concerned exclusively with the adiabatic 
approximation to the dynamics (\ref{action}). This takes place in the
space of holomorphic maps from $\Sigma$ to $\mathbb{CP}^{1}$, i.e.
meromorphic functions on $\Sigma$. They are completely characterised
by the following lemma.

\begin{lemma} \label{lemholom}
Any meromorphic function $W: \Sigma \rightarrow \mathbb{CP}^{1}$ of
degree $n \in \mathbb{Z}$ factorises uniquely as
\[
W=\tilde{W} \circ {\exp},
\]
where $\exp : \Sigma \rightarrow \mathbb{CP}^{1}$ is given by
$\exp (z)=e^{z}$ and $\tilde{W}: \mathbb{CP}^{1} \rightarrow \mathbb{CP}^{1}$
is a rational map of degree $n$.
\end{lemma}

\begin{proof}
Any meromorphic map $W$ on $\Sigma$ can be regarded as a meromorphic map
on $\mathbb{C}$ of period $2\pi i$. We first claim that there is a unique 
meromorphic map $\tilde{W}: \mathbb{CP}^{1}-\{ 0,\infty \} \rightarrow 
\mathbb{CP}^{1}$ such that $W(z)=\tilde{W}(e^{z})$.
This is true because $z \mapsto e^{z}$ is invertible in $\mathbb{C}$ modulo
integer multiples of $2 \pi i$, and this ambiguity does not change the 
value of $W(z)$; that $\tilde{W}$ is meromorphic (and thus a rational map) 
follows from $z \mapsto e^{z}$
being holomorphic and the inverse function theorem in one complex variable.
Since $\exp$ has degree one and the degree is multiplicative with respect
to composition, $\tilde{W}$ has degree $n$.
Finally, $\tilde{W}$ can be extended to a meromorphic map 
$\tilde{W}: \mathbb{CP}^{1} \rightarrow \mathbb{CP}^{1}$ in a
unique way: it cannot have essential singularities at $0$ or $\infty$,
for then the (strong version of the) big Picard theorem (cf.~\cite{Rem}
p.~210) would contradict $n\in \mathbb{Z}$.
\end{proof}

A meromorphic (or antimeromophic) map on $\Sigma$ of degree $n\in \mathbb{Z}$ 
will be called an $n$-{\em lump}. 
Lumps with $n<0$ are sometimes called {\em antilumps}.

\begin{corollary} \label{lemendpts}
For any lump $W: \Sigma \rightarrow \mathbb{CP}^{1}$, the limits
\[
\ell_{\pm}(W):= \lim_{x\rightarrow \pm \infty}W(x+ i y)
\]
are well defined as points of the Riemann sphere.
\end{corollary}

\begin{proof}
The map $\tilde{W}: \mathbb{CP}^{1} \rightarrow \mathbb{CP}^{1}$ 
determined from $W$ by Lemma \ref{lemholom} is continuous,
so this follows from the existence of $\ell_{-}(\exp)=0$
and $\ell_{+}(\exp)=\infty$.
\end{proof}

We shall call $\ell_{-}(W)$ and $\ell_{+}(W)$ the 
{\em endpoints} of $W$. It is easy to
see that the existence of endpoints is a necessary condition for the
Dirichlet energy (\ref{Venergy}) of any map $\Sigma \rightarrow 
\mathbb{CP}^{1}$ to be finite.

\begin{remark} \label{rkpintor}
Lumps with $\ell_{+}(W)=\ell_{-}(W)$ can be interpreted as 
meromorphic functions on the pinched torus depicted in 
Figure~1 --- an elliptic curve with a nodal singularity and 
flat metric (\ref{metricSig}). 
So the results that we shall obtain below for such maps can 
also be interpreted in the context of the 
$\mathbb{CP}^{1}$ $\sigma$-model defined on this singular space.
\end{remark}

\begin{figure}[ht] 
\begin{center}
\vspace{.8cm}
\epsfig{file=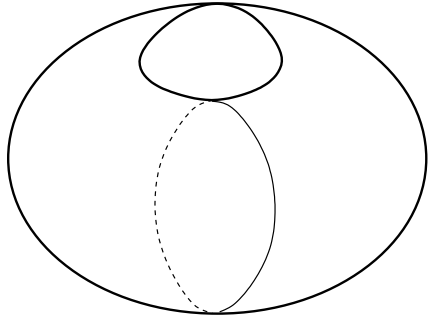}
\small{
\caption{The pinched torus}
}
\end{center}
\label{figpintor}
\end{figure}

%========================================================================

\section{Moduli spaces of lumps} \label{secmsl}

The moduli space of $n$-lumps will be denoted by ${\cal M}_{n}$.
By Lemma~\ref{lemholom}, any $W \in {\cal M}_{n}$ can be written as
\begin{equation} \label{WeqBovA}
W(z)=\frac{B(e^{z})}{A(e^{z})}
\end{equation}
where
\begin{equation} \label{AandB}
A(w)=\sum_{k=0}^{n}c_{k}w^{n-k},
\qquad
B(w)=\sum_{k=0}^{n}c_{n+k+1}w^{n-k}
\end{equation}
are complex polynomials with no common roots, and such that
$c_{n}$ and $c_{2n+1}$ are not both equal to zero. These conditions are
expressed algebraically by the nonvanishing of the resultant of
$A$ and $B$,
\begin{equation} \label{resultant}
\delta(c_{0},\ldots,c_{2n+1}):={\rm Res}(A,B)=
\left|
\begin{array}{ccccccc}
c_{0} & c_{1} & \cdots & c_{n} & & & \\
 & c_{0} & \cdots & c_{n-1} & c_{n} & & \\
 & & \ddots & & & \ddots & \\
 & & & c_{0} & c_{1} & \cdots & c_{n}\\
c_{n+1} & c_{n+2} & \cdots & c_{2n+1} & & & \\
 & c_{n+1} & \cdots & c_{2n} & c_{2n+1} & & \\
 & & \ddots & & & \ddots & \\
 & & & c_{n+1} & c_{n+2} & \cdots & c_{2n+1}\\
\end{array}
\right|. 
\end{equation}
Notice that the polynomials $A$ and $B$ are not uniquely determined from
$W$, but subject to the ambiguity of simultaneous multiplication by an 
element of $\mathbb{C}^{\times}$. Thus we can regard ${\cal M}_{n}$ as
a subset of $\mathbb{CP}^{2n+1}$ through the injection that maps an
$n$-lump $W$
given by (\ref{WeqBovA})--(\ref{AandB}) to the point 
\[
[c_{0}:c_{1}:\cdots : c_{2n+1}] \in \mathbb{CP}^{2n+1}.
\]
The image of this map is the complement of the hypersurface of degree
$n+1$ associated to the homogeneous polynomial $\delta$ in 
(\ref{resultant}), 
\[
{\cal V}(\delta)=\{
[c_{0}:\cdots : c_{n+1}]\in \mathbb{CP}^{n+1} :
\delta(c_{0},\ldots, c_{n+1})=0 \},
\]
and is therefore an open subset in the Zariski topology
of $\mathbb{CP}^{2n+1}$. So we have shown:

\begin{proposition}
${\cal M}_{n}$ is a smooth complex quasiprojective variety of dimension
\[
\dim_{\mathbb{C}}{\cal M}_{n}=2n+1.
\]
\end{proposition}

Let $\mathbb{CP}^{1}_{\Delta}$ denote the diagonal in 
$\mathbb{CP}^{1}\times \mathbb{CP}^{1}$.
For our purposes, it will be useful to give the following description 
of the moduli spaces:

\begin{proposition} \label{propfibration}
There exists a morphism for $n>0$ 
\begin{equation} \label{ellfibration}
\ell: {\cal M}_{n} \rightarrow \mathbb{CP}^{1} \times \mathbb{CP}^{1}
\end{equation}
whereby $\ell=(\ell_{-}, \ell_{+})$ associates to each lump its endpoints;
for $n>1$, this is a fibration by smooth irreducible closed subvarieties of 
${\cal M}_{n}$ with complex dimension $2n-1$.
Moreover, $\ell: {\cal M}_{1} \rightarrow \mathbb{CP}^{1}\times\mathbb{CP}^{1}
- \mathbb{CP}^{1}_{\Delta}$ is an algebraic principal fibre bundle with
structure group $\mathbb{C}^{\times}$.
\end{proposition}

\begin{proof}
We define $\ell$ as the restriction of the rational map 
$\mathbb{CP}^{2n+1} \dashrightarrow \mathbb{CP}^{1} \times \mathbb{CP}^{1}$
given by
\begin{equation} \label{ratmapell}
[c_{0}:\cdots:c_{2n+1}]\mapsto ([c_{n}:c_{2n+1}],
[c_{0}:c_{n+1}]).
\end{equation}
On the complement of the hypersurface ${\cal V}(\delta)$,  
$(c_{n+1}, c_{0})$ and $(c_{2n+1}, c_{n})$ are never equal to 
$(0,0)$; thus (\ref{ratmapell}) is regular on 
${\cal M}_{n}=\mathbb{CP}^{n}-{\cal V}(\delta)$, and $\ell$
is a morphism of algebraic varieties.

The fibre of $\ell$ over $(p,q)=([p_{0}:p_{1}],[q_{0}:q_{1}])$ is obtained 
by intersecting ${\cal M}_{n}$ with the algebraic subset of 
$\mathbb{CP}^{2n+1}$ defined by the homogeneous polynomials
\[
p_{0}c_{2n+1}-p_{1}c_{n}
\qquad
{\rm and}
\qquad
q_{0}c_{n+1}-q_{1}c_{0}.
\]
Since these polynomials are linear in different variables and
nonzero, it is clear that 
\[
\mathbb{C}[c_{0},\ldots,c_{2n+1}]/ \left( p_{0}c_{2n+1}-p_{1}c_{n},
q_{0}c_{n+1}-q_{1}c_{0} \right)
\]
is isomorphic to the ring of complex polynomials in $2n-1$ variables,
whose homogeneous maximal spectrum is $\mathbb{CP}^{2n-1}$. If
$n>1$, it straightforward to verify that this 
$\mathbb{CP}^{2n-1}$ intersects
${\cal V}(\delta)$ transversely independently of $(p,q)$, and
this shows that all the fibres are irreducible, smooth and of dimension
$2n+1$.

For $n=1$, it is easily checked that $\delta$ belongs to the ideal 
\[
\left( p_{0}c_{2n+1}-p_{1}c_{n}, q_{0}c_{n+1}-q_{1}c_{0} \right)
\]
exactly when $p=q$, so that the range of $\ell$ in this case is the
complement of the diagonal $\mathbb{CP}^{1}_{\Delta}$. To show that
$\ell:{\cal M}_{1} \rightarrow \mathbb{CP}^{1}\times\mathbb{CP}^{1}
-\mathbb{CP}^{1}_{\Delta}$ is a principal fibre bundle, we start by
pointing out that the identification of one-lumps with rational maps 
$\mathbb{CP}^{1}\rightarrow \mathbb{CP}^{1}$ of degree one given by 
Lemma~\ref{lemholom} endows ${\cal M}_{1}$ with an algebraic group 
structure, namely
\[
{\cal M}_{1} \cong {\rm PGL}_{2}\mathbb{C}.
\]
More precisely, we identify one-lumps with M\"obius transformations
of $w=e^{z}$. The subgroup generated by rotations and dilations is
isomorphic to $\mathbb{C}^{\times}$, and it is an easy task to verify
that the quotient map
\[ 
{\rm PGL}_{2}\mathbb{C}
\rightarrow
{\rm PGL}_{2}\mathbb{C}
/\mathbb{C}^{\times}
\]
can be identified with $\ell$.
\end{proof}

The pre-image of $(p,q)\in \mathbb{CP}^{1}\times \mathbb{CP}^{1}$
under the map $\ell$ considered in Proposition~\ref{propfibration}
will be denoted by ${\cal M}_{n}^{(p,q)}$; we shall also write
${\cal M}_{n}^{p}:={\cal M}_{n}^{(p,0)}$.
The adiabatic approximation
consists of endowing each of these spaces with a riemannian metric $\gamma$
and studying its geodesic flow, which is a
dynamical system on ${\cal M}_{n}$ by automorphisms of the fibration
$\ell$. The geodesics on the fibres can be 
interpreted physically as a slow motion of lumps of degree $n$ preserving
the endpoints labelling the fibre. Physically, it makes sense to
constrain the motion of the endpoints because we know that in the model
it costs an infinite amount of energy to move them,
which is not available to a lump that starts moving with a finite 
velocity.

The metric $\gamma$ on each ${\cal M}_{n}^{(p,q)}$ is obtained from the 
kinetic energy (\ref{Tenergy}) of the $\sigma$-model. This means that,
if $\zeta_{k}$ ($k=1,\ldots, 2n-1$) are local complex coordinates for 
${\cal M}_{n}^{(p,q)}$,
$\gamma=\gamma_{i\bar{j}}d\zeta_{i}d\bar{\zeta}_{j}$ is defined such that
\begin{equation} \label{gammafromT}
T=\tfrac{1}{2}\gamma_{i\bar{j}}|_{(\zeta_{1}, \ldots,\zeta_{2n-1})}
\dot{\zeta}_{i}\dot{\bar{\zeta}}_{j}.
\end{equation}
Here we allow the $\zeta_{k}$, regarded as parameters
specifying $W$, to depend on time and apply the chain rule.
Geometrically, $\gamma$ can be interpreted as the restriction of the 
$L^{2}$ metric on the infinite-dimensional manifold of smooth maps
$(\Sigma, ds^{2}_{\Sigma}) \rightarrow (\mathbb{CP}^{1}, ds^{2}_{S^{2}})$
to a finite-dimensional submanifold of holomorphic maps with suitable 
boundary conditions. Thus given $W \in {\cal M}_{n}^{(p,q)}$ and two
vectors $X,Y$ of the tangent space
\[
T_{W}{\cal M}_{n}^{(p,q)}=
\{ X \in H^{0}(\Sigma,W^{*}(T^{(1,0)}\mathbb{CP}^{1})) :
\lim_{x\rightarrow -\infty}X|_{x+iy}=0=
\lim_{x\rightarrow +\infty}X|_{x+iy} \},
\]
the metric at $W$ is evaluated as
\begin{equation} \label{gammaasL2}
\gamma|_{W}(X,Y)=\int_{\Sigma}(W^{*}ds^{2}_{S^{2}})(X,Y) \, d\mu_{\Sigma}
\end{equation}
whenever this integral exists.
The main result of this section is the following

\begin{theorem}
The riemannian metric $\gamma$ on ${\cal M}_{n}^{(p,q)}$ relevant 
for the adiabatic approximation is regular for $n\ge 1$. Moreover, it 
is a K\"ahler metric with respect to the complex structure induced by 
${\cal M}_{n}^{(p,q)} \hookrightarrow \mathbb{CP}^{2n+1}$.
\end{theorem}

\begin{proof}
Since we still have the freedom of choosing the inhomogeneous coordinate
on the $\mathbb{CP}^{1}$ target, we may assume without loss of generality 
that $q=0$. This means that we can restrict our attention to maps $W$ 
for which $c_{0}\ne 0$. This condition defines an affine piece of 
$\mathbb{CP}^{2n+1}$ where
\begin{equation} \label{zetas}
\zeta_{k}:=\frac{c_{k}}{c_{0}}, \qquad k=1,\ldots, 2n+1
\end{equation}
are good complex coordinates. Now $q=0$ implies $\zeta_{n+1}=0$ on
${\cal M}_{n}^{p}$. There is one more redundant coordinate on 
${\cal M}_{n}^{p}$ among (\ref{zetas}), and it 
can be eliminated through the equation
\begin{equation} \label{eliminating}
p_{0}\zeta_{2n+1}-p_{1}\zeta_{n}=0,
\end{equation}
where $p=:[p_{0}:p_{1}]$. Suppose first that $p_{0}\ne 0$ holds,
so that $\zeta_{2n+1}$ can be eliminated.
Then a map 
$W \in {\cal M}_{n}^{p}$ can be expressed as
\begin{equation} \label{Winzeta}
W(z)=\frac{\D \sum_{k=1}^{n-1}\zeta_{k+n+1}e^{(n-k)z}
+ p \zeta_{n}}
{\D e^{nz}+ \sum_{k=1}^{n}\zeta_{k}e^{(n-k)z}},
\end{equation}
where we write $p=p_{1}/p_{0}$.
(If $n=1$, the sum in the numerator should be ignored and $p \ne 0$.)
According to (\ref{gammafromT}),
the components of the metric in these coordinates can be read off
as
\begin{eqnarray} 
\gamma_{i\bar{j}}&=&\int_{-\pi}^{\pi}\int_{-\infty}^{+\infty}
\frac{4}{\left(1+|W|^{2}\right)^{2}}
\frac{\partial W}{\partial \zeta_{i}} 
\frac{\partial \bar{W}}{\partial\bar{\zeta}_{j}}
\, dx\, dy \nonumber \\
& = & {2i} \int_{\mathbb{C}}
\frac{1}{(1+|\tilde{W}|^{2})^{2}}
\frac{\partial \tilde{W}}{\partial \zeta_{i}} 
\frac{\partial \bar{\tilde{W}}}{\partial\bar{\zeta}_{j}}
\frac{dw\wedge d\bar{w}}{|w|^{2}} \label{metriccomps}
\end{eqnarray}
where the indices run from $1$ to $2n$ and
we have used the change of variables $w=e^{z}$.  After differentiating 
(\ref{Winzeta}), it is not hard to check that the integrand in
(\ref{metriccomps})
(with respect to the euclidean measure $\frac{i}{2}dw\wedge d\bar{w}$) is a
rational function of $w$ and $\bar{w}$, with the only singularity 
occurring at $w=0$ and being of the form
${\cal O}(|w|^{-1})$ as $|w|\rightarrow 0$,
and with the asymptotic behaviour
${\cal O}(|w|^{-3})$ as $|w|\rightarrow \infty$.
So we conclude that the integral in (\ref{metriccomps}) is finite for
all $i$ and $j$,
which means that the metric $\gamma$ is regular.

To show that the metric is K\"ahler, we start by observing that it
is hermitian with respect to the complex structure associated to the
coordinates $\zeta_{k}$,
\[
\gamma_{i\bar{j}}=\overline{\gamma_{j\bar{i}}}.
\]
The closure of the corresponding $(1,1)$-form can then be seen to be
equivalent to the conditions
\[
\frac{\partial \gamma_{i\bar{j}}}{\partial \zeta_{k}}=
\frac{\partial \gamma_{k\bar{j}}}{\partial \zeta_{i}}, 
\qquad
i,j=1,2,\ldots,n,n+2,\ldots,2n. 
\]
For these to hold, it is sufficient that integration and 
differentiation with respect to $\zeta_{k}$ may be interchanged 
in ${\partial \gamma_{ij}}/{\partial \zeta_{k}}$ 
But this follows from a standard result on Lebesgue integration 
of differentiable maps (cf.\ e.g.\ \cite{LanRFA}, p.~226) once
we observe that the integral
\[
{2i}\int_{\mathbb{C}}
\frac{\partial}{\partial \zeta_{k}}
\left(
\frac{1}{(1+|\tilde{W}|^{2})^{2}}
\frac{\partial \tilde{W}}{\partial \zeta_{i}} 
\frac{\partial \bar{\tilde{W}}}{\partial\bar{\zeta}_{j}}
\right)
\frac{dw\wedge d\bar{w}}{|w|^{2}}
\]
exists and is finite by an argument analogous to the one used to establish
regularity of $\gamma$.

It remains to address the case $p_{0}=0$, i.e.~$p=\infty$. Then
we necessarily have $p_{1}\ne 0$ and (\ref{eliminating}) yields 
$\zeta_{n}=0$. A map $W\in {\cal M}_{n}^{\infty}$ is now expressed as
\begin{equation} \label{Winzetainfty}
W(z)=\frac{\D \sum_{k=1}^{n}\zeta_{k+n+1}e^{(n-k)z}}
{\D e^{nz}+ \sum_{k=1}^{n-1}\zeta_{k}e^{(n-k)z}}
\end{equation}
(where the sum in the denominator should be ignored if $n=1$).
The rest of the argument follows essentially unchanged from 
the case $p_{0}\ne 0$ above.
\end{proof}

%================================================================

\section{Isometries of ${\cal M}_{n}$} \label{secisom}

Our major goal is to compute explicitly the metrics $\gamma$ describing 
the slow motion of lumps in some special situations and interpret their 
geodesics. Not surprisingly, a central part of this study is concerned
with the exploration of isometries, to which we shall now turn. 
In this section, $n$ will not necessarily be taken as nonnegative.

Recall that $\gamma$ is determined from both the metric 
$ds^{2}_{\Sigma}$ on space $\Sigma$ and the metric $ds^{2}_{S^{2}}$ 
on the target, cf.~(\ref{gammaasL2}). These have isometry groups
\begin{equation}\label{semidir}
{\rm Iso}(\Sigma) = V_{4} \ltimes \mathbb{C}^{\times}
\end{equation}
and
\[
{\rm Iso}(S^{2})= {\rm O}(3) \cong \mathbb{Z}_{2} \times \SO(3),
\]
where $V_{4}$ denotes the Vierergruppe $\mathbb{Z}_{2}\oplus
\mathbb{Z}_{2}$. They act on these spaces as follows. The factor
$\mathbb{C}^{\times}$ of ${\rm Iso}(\Sigma)$ refers to the 
translation group of the cylinder,
\[
T_{\lambda }: 
z \mapsto z - \log \lambda, \qquad \lambda \in \mathbb{C}^{\times}, 
\]
whereas the Vierergruppe is generated by any two of the three
transformations
\begin{eqnarray}
& \sigma_{1}: &  z \mapsto -\bar{z}, \label{sig1}\\
& \sigma_{2}: & z \mapsto \bar{z}, \label{sig2}\\
& \sigma_{3}: & z \mapsto -z \label{sig3},
\end{eqnarray}
which also define the semidirect product in (\ref{semidir}).
Notice that both $\sigma_{1}$ and $\sigma_{2}$ reverse
the orientation of $\Sigma$, whereas $\sigma_{3}$ preserves the 
orientation. On the target, the proper rotations in ${\rm SO}(3)$ 
can be represented in terms of M\"obius transformations
of the coordinate $W$ by
\begin{equation} \label{targrot}
R: W \mapsto \frac{\alpha W - \bar{\beta}}{\beta W + \bar{\alpha}},
\qquad |\alpha|^{2} + |\beta|^{2}\ne 0,
\end{equation}
and we can take a reflection across any 
great circle as the generator of the $\mathbb{Z}_{2}$ factor,
say
\[
\sigma: W\mapsto \bar{W}.
\]
It is natural to expect isometries of $({\cal M}_{n}^{(p,q)}, \gamma)$ to 
be produced from the induced action of ${\rm Iso}(\Sigma) \times 
{\rm Iso}(S^{2})$ on $C^{\infty}(\Sigma, \mathbb{CP}^{1})$:
\[
(g,h): W(z)\mapsto h(W(g^{-1}(z))), 
\qquad (g,h) \in {\rm Iso}(\Sigma) \times {\rm Iso}(S^{2}).
\]
In general, these transformations do not preserve the spaces 
${\cal M}_{n}^{(p,q)}$, but it is straightforward to show from the
representation (\ref{gammaasL2}) that they 
still act isometrically as follows:
\begin{eqnarray*} 
 & T_{\lambda}\equiv (T_{\lambda}, \id ): & {\cal M}_{n}^{(p,q)}\rightarrow {\cal M}_{n}^{(p,q)},
\quad \lambda \in \mathbb{C}^{\times}\\
 & \sigma_{1}\equiv (\sigma_{1}, \id): & {\cal M}_{n}^{(p,q)}\rightarrow {\cal M}_{-n}^{(q,p)} \\ 
 & \sigma_{2}\equiv (\sigma_{2}, \id): & {\cal M}_{n}^{(p,q)}\rightarrow {\cal M}_{-n}^{(p,q)} \\ 
 & \sigma_{3}\equiv (\sigma_{3}, \id): & {\cal M}_{n}^{(p,q)}\rightarrow {\cal M}_{n}^{(q,p)}  \\ 
 & R\equiv (\id, R) : & {\cal M}_{n}^{(p,q)}\rightarrow {\cal M}_{n}^{(R(p),R(q))},
\quad R \in {\rm SO}(3) \\ 
 & \sigma \equiv (\id, \sigma): & {\cal M}_{n}^{(p,q)}\rightarrow 
{\cal M}_{-n}^{(\sigma(q),\sigma(p))}. 
\end{eqnarray*}
The next proposition shows how these isometries can be used to
simplify the study of the metrics $\gamma$.

\begin{proposition} \label{propisoms} 
Each fibre of $\ell:{\cal M}_{n}\rightarrow 
\mathbb{CP}^{1}\times \mathbb{CP}^{1}$ is isometric to a fibre of the
form ${\cal M}_{|n|}^{p}$ with $p\in [0,\infty] \subset \mathbb{R}$. 
Moreover, the
isometry groups ${\rm Iso}({\cal M}_{n}^{p})$ of these spaces 
always contain a subgroup isomorphic to
\[
V_{4}\ltimes \mathbb{C}^{\times};
\] 
if $n>1$, ${\rm Iso}({\cal M}_{n}^{p})$ contains a subgroup isomorphic
to $(V_{4}\ltimes \mathbb{C}^{\times})\times{\rm SO}(2)$
if $p=0$ or $p=\infty$.
\end{proposition}
\begin{proof}
Consider the fibre ${\cal M}_{n}^{(p',q')}$ of $\ell$ over arbitrary
$(p',q')$. If $n<0$, we can use $\sigma_{2}$ to map it isometrically
to ${\cal M}_{|n|}^{(p',q')}$. If $q'\ne 0$, we then use the transformation
\[
Q: W \mapsto \frac{W-q'}{1+\bar{q}'W}
\] 
(to be read as $W \mapsto - W^{-1}$ if $q' = \infty$) to map it
to a fibre of the form ${\cal M}_{|n|}^{Q(p')}$. 
Finally, if $Q(p')\not\in [0,\infty]$, we use a rotation 
$W \mapsto e^{i\vartheta} W$ by $\vartheta = -\arg(Q(p'))$.
The composition of these isometries then takes ${\cal M}_{n}^{(p',q')}$
to ${\cal M}_{|n|}^{p}$ for some $p \in [0,\infty]$.

To prove the second part, we start by recalling from above that the 
translations of $\Sigma$ preserve each ${\cal M}_{n}^{p}$, so that
$\mathbb{C}^{\times} \subset {\rm Iso}({\cal M}_{n}^{p})$. This is
not the case for the transformations induced by the generators of
$V_{4}\subset {\rm Iso}(\Sigma)$, 
but they can be combined with target transformations to produce 
$\tilde{\sigma}_{j} \in {\rm Iso}({\cal M}_{n}^{p})$ from the
${\sigma}_{j} \in {\rm Iso}(\Sigma)$ in (\ref{sig1})--(\ref{sig3}). 
Specifically, we take
\begin{eqnarray}
\tilde{\sigma}_{1}& := & R \circ \sigma \circ \sigma_{1} \nonumber ,\\
\tilde{\sigma}_{2}& := & \sigma \circ \sigma_{2}, \label{sigtil2} \\
\tilde{\sigma}_{3}& := & R \circ \sigma_{3}, \label{sigtil3}
\end{eqnarray}
where $R \in {\rm SO}(3)$ is defined by
\begin{equation} \label{rotchanging0p}
R: W \mapsto
\left\{
\begin{array}{c@{,\quad{\rm if}\quad}l}
W & p=0, \\
\D \frac{W-p}{1+pW} & 0 < p < \infty,\\
-W^{-1}& p=\infty.
\end{array}
\right.
\end{equation}
It is clear that the proper rotations of the target giving rise to 
isometries of ${\cal M}_{n}^{p}$
must fix the set $\{ 0,p\}$, so they are either $R$ in (\ref{rotchanging0p})
(leading to $\tilde{\sigma}_{3}$ above) or an element of 
${\rm Stab}_{0}\cap {\rm Stab}_{p}$, and this group is trivial for
$0 < p < \infty$ and ${\rm SO}(2)$ for $p=0$ and $p=\infty$.
However, the case $n=1$ is exceptional: we necessarily have $p\ne 0$, and
in the case of $p=\infty$ target rotations about the endpoints 
act as translations (by an imaginary quantity) on the
whole fibre, so they do not lead to new isometries.
Finally, the target reflections have to be combined
with $\sigma_{1}$ or $\sigma_{2}$ to produce a degree-preserving 
transformation, so no more isometries arise from them.
\end{proof}

The proof above also shows that no further isometries of 
$({\cal M}_{n}^{p},\gamma)$ can be constructed by combining space and
target isometries.

We now state a fundamental lemma relating isometries of a riemannian 
manifold and its totally geodesic submanifolds; these are the 
submanifolds whose geodesics (in the induced metric) are also geodesics 
of the ambient metric (cf.~\cite{CarRG}, p.~132).

\begin{lemma} \label{lemisometries}
Let $S \subset \Iso(M)$ be any set of isometries of a riemannian 
manifold $(M,g)$, and  $F\subset M$ the set of points that are fixed by 
all the elements of $S$. If $F$ is a manifold, it is a totally geodesic
submanifold of $(M,g)$.
\end{lemma}

This elementary result has been used rather crucially in studies of
soliton dynamics, in the case where $F$ is taken to be a finite set
(or the subgroup generated by it), but is also true more generally;
we include a proof in Appendix~A.
The main interest of totally geodesic submanifolds in the context
of soliton dynamics is of course that if their dimension is small 
enough it may be possible to compute the restriction of the relevant 
metric to them.
The geodesics of such manifolds can sometimes be determined (in 
particular, they already are geodesics if their dimension is one),
and they typically describe soliton scattering processes for which the
energy density has some degree of symmetry.
This approach has been exceptionally fruitful in the study of BPS 
monopoles in $\mathbb{R}^{3}$ (see \cite{SutBM} for an 
overview), although in this context it is often more convenient to
impose the relevant symmetries on certain geometric objects parametrised 
by the same moduli spaces as the solutions of the Bogomol'ny\u\i\ equations,
rather than on the metrics directly.

Using the isometries in Proposition~\ref{propisoms} and 
Lemma~\ref{lemisometries}, it is not hard to find 
nontrivial totally geodesic submanifolds for the spaces 
$({\cal M}_{n}^{p},\gamma)$. For instance, if we 
take $S=\{ \tilde{\sigma}_{2} \}$ (cf.~(\ref{sigtil2})) we 
find that $F$ is a subvariety
of real dimension $2n-1$ if $n=2$ and $p\ne 0$, or $n>2$, whereas it
has real dimension $4$ for $n=2$ and $p=0$. Moreover, images of
totally geodesic submanifolds under isometries are again totally
geodesic submanifolds.
A much harder problem is 
to find totally geodesic submanifolds on which the metric and
its geodesics can be 
computed explicitly.
We shall give examples of such in Section~\ref{secsym2l}.

%=======================================================================

\section{Degree-one lumps} \label{secc1l}

For $n=0$, the moduli space is trivially a copy of $\mathbb{CP}^{1}$; 
this follows from Lemma~\ref{lemholom} and the fact that the only 
rational maps of  degree zero are the constants. The map 
$\ell:{\cal M}_{0}\rightarrow \mathbb{CP}^{1} \times \mathbb{CP}^{1}$
analogous to (\ref{ellfibration}) is of course just the embedding of the
diagonal $\mathbb{CP}^{1}_{\Delta}$. 
The adiabatic dynamics as we have defined it in 
Section~\ref{secmsl} is trivial in this degenerate case, because the 
level sets of $\ell$ are either empty or just one point. We regard
${\cal M}_{0}=\mathbb{CP}^{1}$ as a moduli space of classical vacua.

The moduli space of one-lumps is potentially more interesting. Recall that
we established in Proposition~\ref{propfibration} that ${\cal M}_{1}$
has the structure of a principal fibre bundle:
\begin{equation} \label{M1aspfb}
\xymatrix{
\mathbb{C}^{\times} \hookrightarrow {\cal M}_{1} \qquad \quad \ar[d]^\ell \\
  \mathbb{CP}^{1}\times\mathbb{CP}^{1}-\mathbb{CP}^{1}_{\Delta}
}
\end{equation}
It is easy to understand that 
this is just the complexification of the familiar 
description of a two-sphere as a homogeneous space,
\[
S^{2}=\SO (3)/ \SO (2).
\]
The fact that the diagonal $\mathbb{CP}^{1}_{\Delta}$ is absent from the
range of $\ell$ means of course that there is no one-lump $W$ with 
$\ell_{-}(W)=\ell_{+}(W)$; in particular, one-lumps do not exist on a
pinched torus, cf.\ Remark~\ref{rkpintor}. 
This is also a feature of
the $\mathbb{CP}^{1}$ $\sigma$-model on a smooth torus \cite{SpeT2}.

On each fibre ${\cal M}_{1}^{(p,q)}\cong \mathbb{C}^{\times}$ of 
(\ref{M1aspfb}), the structure group acts (transitively and freely) by 
spatial translations, 
which we know to be isometries of the metric 
$\gamma$. It follows from the local isotropy of (\ref{metricSig}) that 
the metric on ${\cal M}_{1}^{(p,q)}$ is completely described by a 
constant $m(p,q)$.
By fixing a one-lump $W_{0} \in {\cal M}_{1}^{(p,q)}$, we can introduce
a global coordinate 
$\zeta\in \mathbb{C}^{\times} \cong \Sigma$ and parametrise any
other  $W_{0} \in {\cal M}_{1}^{(p,q)}$ as 
$W(z)=W_{0}(z-\zeta)$; if we define $c$ locally
by $c:=\log \zeta$, we can write down the metric conveniently as
\[ 
\gamma= m(p,q) \, dc\, d\bar{c}.
\]
This constant $m(p,q)$ can be interpreted as the mass of a lump with 
endpoints $p$ and $q$, and adiabatic motion in ${\cal M}_{1}^{(p,q)}$ 
is just rigid motion on $\Sigma$ with inertia given by this constant.

It is straightforward to verify that two one-lumps with endpoints 
$p_{1}, q_{1}$ and $p_{2}, q_{2}$ at 
the same distance $d(p_{1},q_{1})=d(p_{2},q_{2})$ have the same 
shape, in the sense that their energy 
densities 
${\cal E} \in L^{2}(\Sigma, d\mu_{\Sigma})$, given by
\begin{equation} \label{endens}
{\cal E}(z)=\frac{4 |\partial_{z}W(z)|^{2}}
{\left(1+|W(z)|^{2} \right)^{2}}
\end{equation}
are related by a spatial translation. The possible shapes of one-lumps
are classified through $d$ by the points of the interval $]0,\pi]$. It 
follows from these observations that on each ${\cal M}_{1}^{p,q}$ a 
one-lump moves without altering its shape, and that $m(p,q)$ can be 
expressed as a function of $d(p,q)$.
In Figure~2, we plot the energy density profiles of one-lumps
with different shapes. For $d=\pi$, the lump profile has circular 
symmetry, which is hardly surprising -- it is readily checked that
(\ref{endens}) is invariant under any global target rotation (\ref{targrot}),
and we know from the proof of Proposition~\ref{propisoms} that for $n=1$
and antipodal endpoints a translation of $z$ by an imaginary quantity is
equivalent to a target rotation about the endpoints.
When we decrease $d$, the profile acquires a peak, which becomes more and 
more pronounced as the endpoints approach each other.

\begin{figure} \label{figshapes}
\begin{center}
\begin{tabular}{ccc}
\epsfig{file=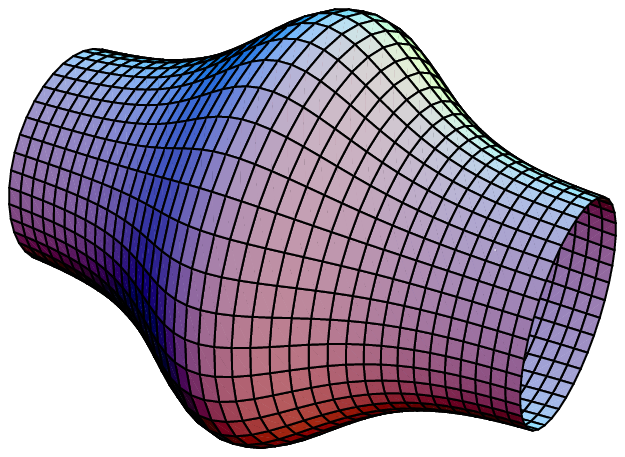,width=4.5cm}
\hspace{.3cm} &
\raisebox{.2cm}
{\epsfig{file=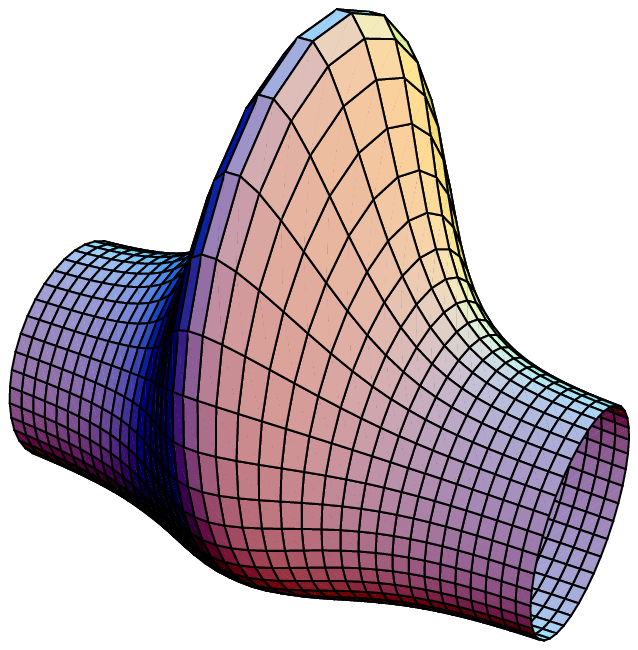,width=4.5cm}}
\hspace{.3cm} &
\raisebox{.8cm}
{\epsfig{file=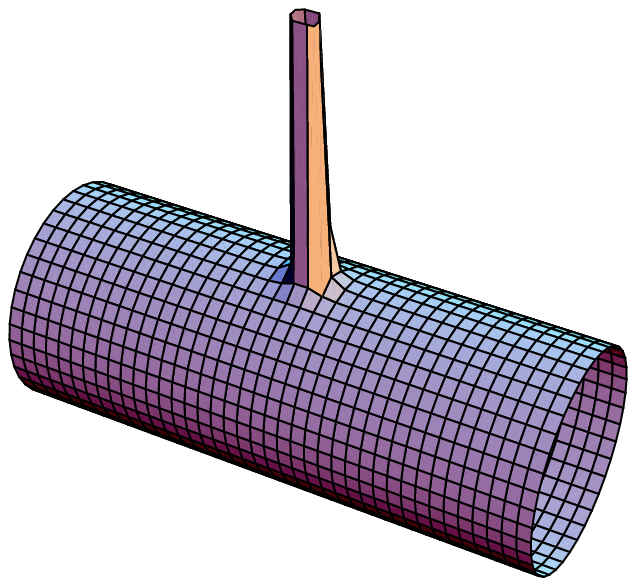,width=3.7cm}} \\
(a) $d=\pi$ & (b) $d=2\pi/3$ & (c) $d=\pi/250$
\end{tabular}
\end{center}
\vspace{.5cm}
\caption{\small \em
One-lumps with different shapes: 
{\em (a)} $W(z)=e^{-z}$; 
{\em (b)} $W(z)=\sqrt{3}/(2ie^{z}+1)$;
{\em (c)} $W(z)=\tan(\tfrac{\pi}{500})/(i e^{z}+1)$.
The energy density profiles are plotted radially on top of a cylinder
of unit radius for $|\Rea z|\le 5/2$.
}
\end{figure} 

In fact, we can show that even one-lumps of different shapes have the 
same mass in the adiabatic approximation. This is just a special case of 
the following fact:

\begin{proposition} \label{propm1}
The mass of any $n$-lump is $4 \pi n$. In particular, the $L^2$ metric on
${\cal M}_{1}^{(p,q)}$ is $\gamma=4 \pi \, dc\,d\bar{c}$.
\end{proposition}
\begin{proof}
Let $W$ be any $n$-lump and introduce a coordinate $c=\log \zeta$,
$\zeta \in \mathbb{C}^{*}$ as above. 
According to (\ref{metriccomps}), the mass of $W$ is then given by the
integral
\begin{eqnarray*}
m & = & 
\int_{\Sigma} 
\frac{4}{\left( 1+|W(z)|^2 \right)^2}
\left| \left. \frac{\partial}{\partial c}\right|_{c=0}W(z-c) \right|^2
{d\mu_{\Sigma}}\\
& = & 2i
\int_{\Sigma} 
\frac{4}{\left( 1+|W(z)|^2 \right)^2}
\left( \left| \frac{\partial W}{\partial z}(z) \right|^2+
\left| \frac{\partial W}{\partial \bar{z}}(z) \right|^2 \right)
{d\mu_{\Sigma}}\\
& = & V[W] = 4 \pi n,
\end{eqnarray*}
where we made use of the Bogomol'ny\u\i\ equation (\ref{CReq}) to complete 
the potential energy~(\ref{Venergy}). 
\end{proof}

%===================================================================

\section{Incompleteness of multilump metrics} \label{secincomp}

As we have seen, the adiabatic dynamics of one-lumps can be described as 
constant
motion on the translation group of $\Sigma$; the shape of the one-lump
is fixed by the initial endpoints and the dynamical moduli can be interpreted
as a centre of mass. For $n \ge 2$ however, typical
dynamical processes will include relative motion determined by the 
interactions among the individual solitons entwined within a given 
field configuration, and the metrics are correspondingly more complicated. 
In this section, we shall establish an important property of the multilump 
metrics, which accounts for the possiblity of lump collapse in finite time:

\begin{theorem} \label{thmincomplete}
For $n \ge 2$, the $L^{2}$ metric (\ref{gammaasL2}) on 
${\cal M}_{n}^{(p,q)}$ is incomplete for any 
$(p,q)\in \mathbb{CP}^{1}\times\mathbb{CP}^{1}$.
\end{theorem}
\begin{proof}
By Proposition~\ref{propisoms}, we do not loose generality by taking 
$p\in [0,\infty]$ and $q=0$. Our strategy will
be to exhibit (for each $n\ge 2$) particular paths 
$\gamma_{p}:[a,b[\,\subset \mathbb{R} \rightarrow {\cal M}_{n}^{p}$
such that:
\begin{itemize}
\item
$\D \lim_{t\rightarrow b} \gamma_{p}(t) \notin {\cal M}_{n}^{p}$;
\item
$\gamma_{p}$ has finite length in the metric (\ref{gammaasL2}).
\end{itemize}
According to (\ref{metriccomps}), the length of $\gamma_p$ is given by
\begin{equation} \label{length}
L(\gamma_{p})=\int_{a}^{b}\left(
\int_{\mathbb{C}}\Phi(w,t)\,d^{2}w 
\right)^{1/2}dt,
\end{equation}
where
\begin{equation} \label{Phi}
\Phi(w,t)=\frac{4}{(1+|\tilde{W}_{t}|^{2})^{2}}
\left| \frac{\partial \tilde{W}_{t}}{\partial t} \right|^{2} 
\frac{1}{|w|^{2}} 
\end{equation}
and we denote by $w\mapsto \tilde{W}_{t}(w)$ the rational map
$\mathbb{CP}^{1}\rightarrow \mathbb{CP}^{1}$ corresponding to
$\gamma_{p}(t)$.
It is convenient to consider three separate cases: 
(i) $p=0$, 
(ii) $0<p<\infty$ and
(iii) $p=\infty$. 
In each case, we choose $\gamma_{p}$ having future
convenience in mind.\\[.2cm]
\noindent 
(i) $p=0:$ \\[.2cm]
We define $\gamma_{0}$ on $[1,+ \infty[$ by
\begin{equation}\label{gamma0}
\gamma_{0}(t):w\mapsto \frac{2w}{t(w^{n}+1)}=\tilde{W}_{t}(w).
\end{equation}
Notice that, for $t\in [1,+\infty[$, $\tilde{W}_{t}$ is a rational map of
degree $n$ with the required boundary conditions 
$\tilde{W}_{t}(0)=0=\tilde{W}_{t}(\infty)$, thus $\gamma_{0}$ is well defined.
Moreover, $\tilde{W}_{\infty}:=\lim_{t\rightarrow +\infty}\tilde{W}_{t}$
is not a rational map:
\[
\tilde{W}_{\infty}(w)=
\left\{
\begin{array}{c@{,\quad{\rm if}\quad}l}
0 & w^{n}\neq -1 \\
\infty & w^{n}=-1 ;
\end{array}
\right.
\]
so indeed $\gamma_{0}(t)$ leaves ${\cal M}_{n}^{0}$ as 
$t\rightarrow +\infty.$

We now set to prove that $\gamma_{0}$ has finite length. This is
given by (\ref{length}) with
\[
\Phi(w,t)= \frac{4}{(t^{2}+|\tilde{W}_{1}|^{2})^{2}}
\frac{|\tilde{W}_{1}|^2}{|w|^2}.
\]
We denote by $\{ w_{j} \}_{j=1}^{n}$ the set of $n$th roots of $-1$, which
are all the (simple) poles of $\tilde{W}_{1}$, and fix $\varepsilon \in 
\,]0,\frac{1}{2}\sin\frac{\pi}{n}[$. It is now convenient to write 
\[
\int_{\mathbb{C}}\Phi(\cdot,t)=
\left( 
\int_{B_{1/\varepsilon}(0)-\cup_{j=1}^{\infty}B_{\varepsilon}(w_{j})}+
\int_{\mathbb{C}-B_{1/\varepsilon}(0)}+
\sum_{j=1}^{n}\int_{B_{\varepsilon}(w_{j})}
\right)\Phi(\cdot,t)
\]
and estimate each of the integrals separately.
In the following, $C, C_{0},$ etc.~denote positive constants (dependent
on $\varepsilon$ only).

Since $w\mapsto \tilde{W}_{1}(w)/w $ has modulus bounded on
$B_{1/\varepsilon}(0)-\cup_{j=1}^{\infty}B_{\varepsilon}(w_{j})$ 
(say by a constant $C_{0}$) independently of $t$, we may write
\[
\int_{B_{1/\varepsilon}(0)-\cup_{j=1}^{\infty}B_{\varepsilon}(w_{j})}
\Phi(\cdot,t)  < 
\int_{B_{1/\varepsilon}(0)-\cup_{j=1}^{\infty}B_{\varepsilon}(w_{j})}
\frac{4 |\tilde{W}_{1}|^{2}d^2 w}{t^{4}|w|^2}
 <   \int_{B_{1/\varepsilon}(0)}\frac{4 C_{0}^{2}d^2 w}{t^{4}}=
\frac{4 \pi C_{0}^{2}}{\varepsilon^{2} t^{4}}.
\]
Similarly, $w\mapsto w \,\tilde{W}_{1}(w)$ is also bounded in modulus for 
$|w|>1/\varepsilon$ (by $C_{\infty}$, say), hence
\[
\int_{\mathbb{C}-B_{1/\varepsilon}(0)}\Phi(\cdot,t) <
\int_{\mathbb{C}-B_{1/\varepsilon}(0)}
\frac{4 |\tilde{W}_{1}|^{2}d^2 w}{t^{4}|w|^2}
< \int_{\mathbb{C}-B_{1/\varepsilon}(0)}\frac{4 C_{\infty}^{2} d^2 w}
{t^4 |w|^4}
= \frac{4 \pi C_{\infty}^{2}}{\varepsilon^{2} t^{4}}.
\]

On $B_{\varepsilon}(w_{j})$, the function 
$w \mapsto (w-w_{j})\tilde{W}(w)$ has no poles or zeroes,
so there is a constant $C_{j}>1$ satisfying
\[
C_{j}^{-1} < |(w-w_{j})\tilde{W}(w)| < C_{j}.
\]
Therefore,
\[
\int_{B_{\varepsilon}(w_{j})}\Phi(\cdot, t)<
\frac{4}{(1-\varepsilon)^{2}}
\int_{B_{\varepsilon}(w_j)}
\frac{\frac{C_{j}^{2}}{|w-w_j|^2}}
{\left( t^2 + \frac{C_{j}^{-2}}{|w-w_j|^2}\right)^{2}} \;\; d^2 w.
\]
We now make the change of variable $w \mapsto v:=tC_{j}(w-w_{j})$
and estimate the right-hand-side of the inequality above as follows:
\begin{eqnarray*}
\frac{4 C_{j}^{2}}{(1-\varepsilon)^2 t^{4}}\int_{B_{t\varepsilon C_{j}}(0)}
\frac{|v|^{2} \,d^2 v}{(|v|^{2}+1)^2} & = &
\frac{8\pi C_{j}^{2}}{(1-\varepsilon)^2 t^{4}}
\int_{0}^{t\varepsilon C_{j}}\frac{|v|^3 \,d|v|}{(|v|^2 +1)^{2}}\\
& = & \frac{8\pi C_{j}^{2}}{(1-\varepsilon)^2 t^{4}}
\int_{0}^{1+(t\varepsilon C_{j})^2}\frac{u-1}{u^2}\,du \\
& < & \frac{8\pi C_{j}^{2}}{(1-\varepsilon)^2 t^{4}}
\log(1+(t\varepsilon C_{j})^2).
\end{eqnarray*}
There are constants $C'_{j}$ and $C''_{j}$ such that the last 
expression above is bounded by
\[
{C'_{j}}{t^{-4}}+{C''_{j}}{t^{-4}}\log t.
\]
Putting together all the estimates above, we conclude that there is an
overall constant $C$ such that the length of $\gamma_{0}$ satisfies the
inequality
\[
L(\gamma_{0})< C \int_{1}^{\infty}\frac{1}{t^{2}}\sqrt{1+ \log t} \, dt,
\]
in which the right-hand-side is finite.

Before we proceed, we would like to remark that we defined $\gamma_{0}$
as a target scaling of $W_{1}\in{\cal M}_{n}^{0}$ in (\ref{gamma0}), and 
that any other choice of $W_{1}$ would lead (through scaling) to a path in 
${\cal M}_{n}^{0}$ that would suit our purposes (cf.~\cite{SadSpe}). 
However, the scaling of
a given map within a fibre ${\cal M}_{n}^{p}$ does {not} yield paths 
of finite length if $p\neq 0$.\\[.2cm]
\noindent 
(ii) $0<p<\infty:$ \\[.2cm]
We now take $\gamma_{p}$ with domain $[\frac{1}{2},1[$ and defined by
\[
\gamma_{p}(t):w\mapsto \frac{tp(tw+1)}{(1-w)^{n-1}(w+t)}=\tilde{W}_{t}(w).
\]
It is easy to check that $W_{t} \in {\cal M}_{n}^{p}$ for each
$t\in [\frac{1}{2},1[$. Now $W_{1} \notin {\cal M}_{n}^{p}$ because
$\tilde{W}_{1}$ is a map of degree $n-1$.

The length of $\gamma_{p}$ is given by (\ref{length}), with
\[
\Phi(w,t)=\frac{4p^2 |2tw+t^2 + 1|^2 |w-1|^{2n-2}}
{(|w-1|^{2n-2}|w+t|^2 + |tp|^2 |tw+1|^2)^2}.
\]
We fix now $\varepsilon \in \,]0,\frac{1}{2}[$ and write
\[
\int_{\mathbb{C}}\Phi(\cdot,t)=
\left( \int_{B_{\varepsilon}(0)} + 
\int_{\mathbb{C}-(B_{\varepsilon}(0)\cup B_{\varepsilon}(-1))}
+\int_{B_{\varepsilon}(-1)}\right)\Phi(\cdot,t).
\]
The first integrals do not cause problems, as we can write for suitable
constants $C_{0}$ and $C_{\infty}$ and all $t\in [\frac{1}{2}, 1[$
\[
\int_{B_{\varepsilon}(0)}\Phi(\cdot,t)< \int_{B_{\varepsilon}(0)}
C_{0}\,d^2 w=
\pi\varepsilon^2 C_{0}
\]
and
\[
\int_{\mathbb{C}-(B_{\varepsilon}(0)\cup B_{\varepsilon}(-1))}
\Phi(\cdot,t)< 
\int_{\mathbb{C}-B_{\varepsilon}(0)}\frac{C_{\infty}\,d^2 w}{|w|^{2n}}=
\frac{\pi C_{\infty}}{(n-1)\varepsilon^{2n-2}}.
\]
As $t\rightarrow 1$ however, the integral over $B_{\varepsilon}(-1)$
becomes unbounded, but we shall show that the lenght $L(\gamma_{p})$ 
remains finite. We change variables as $w \mapsto v:=(1-t)^{-1}(w+1)$
and estimate
\begin{eqnarray*}
\int_{B_{\varepsilon}(-1)}\Phi(\cdot,t)
& < & 
\int_{B_{\frac{\varepsilon}{(1-t)}}(0)} \frac{C_{1}|(1-t)+2tv|^{2}}
{(|v+1|^2 + C_{2}|tv+1|^2)^{2}} \;d^2 v\\
& < &{C_{1}}\int_{B_{\frac{\varepsilon}{(1-t)}}(0)}
\frac{(1-t)^2 + 4(1-t)t|v|+4t^2 |v|^2}{(|v+1|^2+C_{2}|tv+1|^2)^2}\; d^2 v\\
& < & {C_{3}}\int_{B_{\frac{\varepsilon}{(1-t)}}(0)}
\frac{(1-t)^2 + 4(1-t)t|v|+4t^2 |v|^2}{(|v|^2+1)^2} \, d^2 v\\
& = & {2 \pi C_{3}}\int_{0}^{\frac{\varepsilon}{(1-t)}}
\frac{(1-t)^2|v| + 4(1-t)t|v|^2+4t^2 |v|^3}{(|v|^2+1)^2} d|v|\\
& = & {2 \pi C_{3}}\left[
\frac{(1-t)^2}{2}\frac{\varepsilon^{2}}{\varepsilon^2 + (1-t)^2}\right.\\
& & + \, 2(1-t)t\left( \arctan \frac{\varepsilon}{1-t} 
-  \frac{\varepsilon (1-t)}{\varepsilon^2 + (1-t)^2}\right) \\ 
& & \left. +\, 2t^{2}\left( \log \frac{\varepsilon^2 + (1-t)^2}{(1-t)^2}
-\frac{\varepsilon^2}{\varepsilon^2 + (1-t)^2}\right)
\right]\\
& < & {2 \pi C_{3}} \log\left(1+\frac{\varepsilon^2}{(1-t)^2}
\right)\\
& < & {4 \pi C_{3}}\log\frac{1}{1-t},
\end{eqnarray*}
where again $C_{1}$, $C_{2}$ and $C_{3}$ are suitable constants
dependent on $\varepsilon$ only. Hence the length of 
$\gamma_{p}$ is bounded above by a finite quantity:
\begin{equation}\label{boundLgp}
L(\gamma_{p})<C\int_{1/2}^{1}
\left(1+\log \frac{1}{1-t} \right)^{1/2} dt.
\end{equation} \\[.2cm]
\noindent 
(iii) $p=\infty:$ \\[.2cm]
Finally, we define the path $\gamma_{\infty}$ with domain $[\frac{1}{2},1[$
by
\[
\gamma_{\infty}(t):w\mapsto \frac{tw+1}{w^{n-1}(w+t)}=\tilde{W}_{t}(w).
\]
Again, it is easy to check that this defines a path on the fibre 
${\cal M}_{n}^{\infty}$ with $\tilde{W}_{1}$ having degree $n-1$.
In the formula (\ref{length}) for the length of $\gamma_{\infty}$ we
now have
\[
\Phi(w,t)=\frac{4|w-w^{-1}|^2}
{(|w|^{n-1}|w+t|^2 + |tw+1|^2 |w|^{-n+1})^2}.
\]
The rest of the argument is completely analogous (if somewhat easier) to 
case (ii) above, and we are again led to a finite bound for $L(\gamma_\infty)$ 
identical to (\ref{boundLgp}).
\end{proof}

%===================================================================

\section{Dynamics of degree-two lumps} \label{secdc2l}

To have some insight on the nontrivial scattering of multilumps, and in 
particular how the lump collapse established in Theorem~\ref{thmincomplete} 
may be realised, we may hope to compute particular geodesics of the multilump 
metrics. This is a very difficult task, but in this section we show that
some results can be obtained in the simplest case of $n=2$ lumps.

\subsection{Symmetric two-lumps} \label{secsym2l}

Even for $n=2$, $\dim_{\mathbb{R}}{\cal M}_{2}^{p}=6$ is too large
to render the computation of the total metric feasible. We start by
restricting our attention to a totally geodesic submanifold.

\begin{lemma} \label{lemtotalgeod}
The following are totally geodesic submanifolds for the metrics 
$\gamma$:
\begin{enumerate}
\renewcommand{\labelenumi}{(\roman{enumi})}
\item
$\tilde{\Xi}_{0}:= 
\{z \mapsto \frac{\alpha}{\cosh z + \beta} : \alpha \in \mathbb{C}^{\times},
\beta \in \mathbb{C}
\} \subset {\cal M}_{2}^{0}$
\item
${\Xi}_{\infty}:=
\{ z\mapsto \frac{e^{-z}+ \alpha}{e^{z} + \alpha} : \alpha\in \mathbb{C}
-\{ -1, 1\}\} \subset {\cal M}_{2}^{\infty}$.
\end{enumerate}
\end{lemma}

\begin{proof}
Part (i) follows from the direct application of 
Lemma~\ref{lemisometries} to the
set $S$ consisting of the isometry $\tilde{\sigma}_{3}$ defined in 
(\ref{sigtil3}), and using the parametrisation (\ref{Winzeta}) for
$W\in{\cal M}_{2}^{p}$. Part (ii) follows from the
same argument (now using (\ref{Winzetainfty}) to express $W$), 
combined with the application of isometries of the
form $W(z)\mapsto - W(z)$ and $W(z)\mapsto W(-z)$ discussed in 
Proposition~\ref{propisoms}. 
\end{proof}

We shall now focus on the two cases $p=0$ and $p=\infty$ separately.

\subsubsection{$p=0$} \label{secs2l0}

The submanifold $\tilde{\Xi}_{0}$ in Lemma~\ref{lemtotalgeod} 
has real dimension
four. Computing the restriction of the metric to it is still too
complicated, but we can achieve this in certain submanifolds of 
codimension two. If they are totally geodesic 
in $\tilde{\Xi}_{0}$, they will also be in $({\cal M}_{2}^{0},\gamma)$.

We start by applying again Lemma~\ref{lemisometries} to 
$\tilde{\Xi}_{0}$ with $S$ consisting of the isometry
\[
W(z) \mapsto - W(z-i\pi).
\]
The fixed point set in $\tilde{\Xi}_{0}$ is
\[
\Xi_{0}:=\{ z \mapsto \alpha \sech z : \alpha \in 
\mathbb{C}^{\times}\}\subset {\cal M}_{2}^{0},
\]
which is a two-dimensional totally geodesic
submanifold. The $\SO(2)$ isometry subgroup
of target rotations acts on $\Xi_{0}$, and this implies that the restriction
of the metric to this submanifold is independent of 
$\vartheta:=\arg{\alpha}$. It can be written as
\begin{equation} \label{metric00}
\gamma|_{\Xi_{0}}=I(a)
(da^{2}+a^{2}d\vartheta^{2})
\end{equation}
where $a:=|\alpha|$ and $I(a)$ is given from (\ref{metriccomps}) as
\begin{equation} \label{intcsi0}
I(a)=4\int_{-\pi}^{\pi}\int_{-\infty}^{\infty}
\frac{|\sech z|^{2}}{(1+a^{2}|\sech z|^{2})^{2}}\, dx\, dy.
\end{equation}

We shall now explain how to compute the integral (\ref{intcsi0}) in 
closed form. We introduce the quantity
\[
J_{\Lambda}(a):=
2i \int_{\Sigma} \left(
\frac{1}{1+a^{2}|\sech z|^{2}}-\Lambda(|\sech z|^{2}) 
\right) dz \wedge d\bar{z}
\]
dependent on a {\em regulator} $\Lambda$, which we define to be an 
integrable function $\Lambda:[0,+\infty[\rightarrow 
\mathbb{R}$ with $\supp(\Lambda)\subset [0,1]$ such that 
$J_{\Lambda}(a)$ above exists as a real number. 
Our aim will be to 
compute $J_{\Lambda}(a)$ for
suitable $\Lambda$, and then determine $I(a)$ as
\begin{equation} \label{IfromJ}
I(a)=-\frac{1}{2a} \frac{d}{da}J_{\Lambda}(a).
\end{equation}
It should be noted that the value of $I(a)$ is then independent of the
regulator; more precisely, it will become clear (cf.~(\ref{Jfinal}) below) 
that $I(a)$ as given by (\ref{IfromJ}) is
invariant under any of the transformations
\[
\Lambda(r)\mapsto \Lambda(r) + r \lambda(r)
\]
where $\lambda$ is an element of $L^{2}([0,+\infty[)$
supported on a subset of $[0,1]$.

To calculate $J_{\Lambda}(a)$, we start by changing variables using
$z \mapsto u=\sech^{2}z$; this is a map $\Sigma \rightarrow \mathbb{CP}^{1}$ 
of degree four, so we obtain
\[
J_{\Lambda}(a)={2i}\int_{\mathbb{C}}\left(
\frac{1}{1+a^{2}|u|}-\Lambda(|u|)
\right) \frac{du \wedge d\bar{u}}{|u|^{2}|u-1|}.
\]
In terms of polar coordinates $r$ and $\theta$ for the $u$-plane,
\begin{eqnarray*}
J_{\Lambda}(a) & = & \D
4 \int_{0}^{\infty}\left( 
\int_{0}^{\pi}\frac{d \theta}{\sqrt{r^{2}+1-2r \cos \theta}}+
\int_{0}^{\pi}\frac{d \theta}{\sqrt{r^{2}+1+2r \cos \theta}}
\right)
\left(
\frac{1}{1+a^{2}r}-{\Lambda(r)}
\right) \frac{dr}{r} \\
& = & \D 16 \int_{0}^{\infty}\left(
\frac{1}{1+a^{2}r}-{\Lambda(r)}
\right)
\frac{K(k(r)) dr}{r(r+1)}.
\end{eqnarray*}
Here, $K$ is Legendre's complete elliptic integral of the first 
kind, and we have made use of the standard formulas (289.00) and (291.00) in 
\cite{ByrFri}, with
\begin{equation} \label{kofr}
k(r)=\frac{2\sqrt{r}}{r+1}.
\end{equation}

To proceed, we change the variable of integration from $r$ to
\[
c:=\frac{1-k'}{1+k'},
\]
whereby $k':=\sqrt{1-k^{2}}$ as usual. This requires some care, since
(\ref{kofr}) is not injective, but can be inverted as
\begin{eqnarray*}
\D r=\frac{1-\sqrt{1-k^{2}}}{1+\sqrt{1-k^{2}}}={c} 
& {\rm for} & r \in [0,1), \\
\D r=\frac{1+\sqrt{1-k^{2}}}{1-\sqrt{1-k^{2}}}=\frac{1}{c} 
& {\rm for} & r \in [1,\infty).
\end{eqnarray*}
By making use of Landen's transformation (cf.\ e.g.~\cite{Tri}, p. 238)
\[
K(k)=\frac{2}{1+k'}K\left(\frac{1-k'}{1+k'}\right),
\]
we then arrive at
\begin{equation}\label{Jfinal}
J_{\Lambda}(a)= 16 \int_{0}^{1} \left(
\frac{c}{c+a^{2}}-\frac{a^{2}}{1+a^{2}c}+\frac{1-\Lambda(c)}{c}
\right) K(c) dc.
\end{equation}

At this stage, we choose the particular regulator
\[
\Lambda(c)=
\left\{
\begin{array}{c@{,\quad{\rm if}\quad}l}
c+1 & 0\le c \le 1, \\
0 & c> 1,
\end{array}
\right.
\]
and drop the $\Lambda$ subscript in $J_{\Lambda}$ to obtain
\begin{equation} \label{regJ}
J(a)=- 16 a^2 \int_{0}^{1}
\left(
\frac{1}{c+a^{2}} +
\frac{1}{1+a^{2}c}
\right) K(c) dc.
\end{equation}
The integral above can be evaluated
in closed form by making use of the following result, which we
prove in Appendix~B:

\begin{lemma} \label{lemintegr}
The integral
\begin{equation} \label{functionf}
f(t):=\int_{0}^{1}\left( 
\frac{1}{k+t}+\frac{1}{1+tk}
\right)K(k)dk
\end{equation}
defines an analytic function on $]0,+\infty[$ which satisfies
\begin{equation} \label{closedf}
f(t)=  
\left\{
\begin{array}{c@{,\quad{\rm if}\quad}l}
\D \frac{\pi}{2} K\left(\sqrt{1-t^{2}}\right) & 0 < t \le 1, \\[.5cm]
\D \frac{\pi}{2t} K\left(\sqrt{1-t^{-2}}\right) & t > 1.
\end{array}
\right.
\end{equation}
\end{lemma}

Thus we can write (\ref{regJ}) as
\[
J(a)=-16 a^{2} f(a^{2})
\]
and determine from (\ref{IfromJ}) the conformal factor in the metric 
(\ref{metric00}) as
\begin{eqnarray} 
I(a)& = & 16 f(a^{2}) + 16 a^{2}f'(a^{2}) \nonumber \\
& = &
\left\{
\begin{array}{c@{,\quad{\rm if}\quad}l}
\D \frac{8 \pi}{a^{4}-1}
\left( E\left( \sqrt{1-a^{4}}\right)-K\left( \sqrt{1-a^{4}}\right)\right)
 & 0<a\le 1, \\[.5cm]
\D\frac{8 \pi}{a^{4}-1}
\left( a^{2} E\left( \sqrt{1-a^{-4}}\right)-a^{-2} K\left( 
\sqrt{1-a^{-4}}\right)\right) 
& a > 1,
\end{array}
\right. \label{Iofa}
\end{eqnarray}
where $E$ is Legendre's complete elliptic integral of the second kind;
here, we made use of 
\[
\frac{dK}{dk}=\frac{E(k)-(1-k^{2})K(k)}{k(1-k^{2})}.
\]
Notice that the function $I$ is smooth on $]0,\infty]$; we plot a section
of its graph in Figure~3.

\begin{figure} \label{figIvsa}
\begin{center}
\vspace{1cm}
\epsfig{file=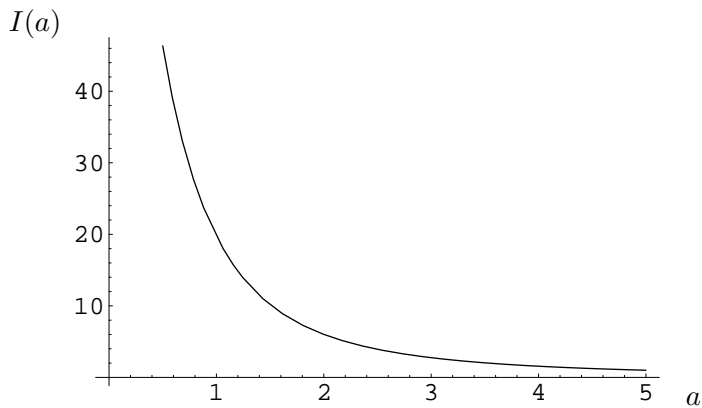,width=8cm}\\
\vspace{-5.5cm}\hspace{-8.75cm}$I(a)$\\
\vspace{4.5cm}\hspace{8.75cm}$a$\\
\vspace{.5cm}
\caption{\small \em
The conformal factor $I(a)$ for the metric on $\Xi_{0}$.
}
\end{center}
\end{figure}

We now show that the conformal factor $I(a)$ given by (\ref{Iofa}) is such
that $\Xi_{0}$ can be embedded in euclidean $\mathbb{R}^{3}$:

\begin{lemma} \label{lemembed}
The riemannian manifold $\Xi_{0}$ with metric given by 
equations (\ref{metric00}) and
(\ref{Iofa}) can be isometrically embedded in $\mathbb{R}^{3}$ as a surface 
of revolution.
\end{lemma}
\begin{proof}
A general surface of revolution in $\mathbb{R}^{3}$ is described by 
an embedding (in cartesian coordinates)
\[
(a, \vartheta) \mapsto (a\, u(a) \cos \vartheta, a\,u(a) \sin \vartheta, 
v(a) ),
\]
where $\vartheta$ is a standard local coordinate on the circle and we take 
$a>0$. Under this map, the euclidean metric of $\mathbb{R}^{3}$ pulls back as
\[
\left( \left(u(a)+a u'(a)\right)^{2}+ v'(a)^{2} \right) da^{2}+
a^{2} u(a)^{2} d\vartheta^{2}.
\]
In order for this to be the polar isothermal form (as in equation 
(\ref{metric00})) of a metric in two dimensions, one must set the
coefficient of $da^{2}$ equal to $u(a)^{2}$, yielding
\begin{equation} \label{vofa}
v(a)=\int^{a}\sqrt{-s^{2}u'(s)^{2}-
s(u(s)^{2})'} ds + {\rm const.}
\end{equation}
This determines a real function if and only if the condition
\begin{equation} \label{embexists}
a (u'(a))^{2}\le (u(a)^{2})'
\end{equation}
is satisfied for all $a$. In our case of interest, we should take
\[
u(a)=\sqrt{I(a)}.
\]
It is easy to check that $u'(a)<0$ for all $a>0$ and (\ref{embexists})
can be expressed as
\[
\frac{d}{da}\log I(a) \ge -\frac{4}{a},
\] 
which can be verified to hold for $a\in \,]0,+\infty[$.
\end{proof}

The embedded surface has cylindrical topology and provides a good picture 
of the geometry of $\Xi_{0}$;
we plot a section of it in Figure~4, using the construction in 
Lemma~\ref{lemembed}. 
\begin{figure} \label{figembsurf}
\begin{center}
\epsfig{file=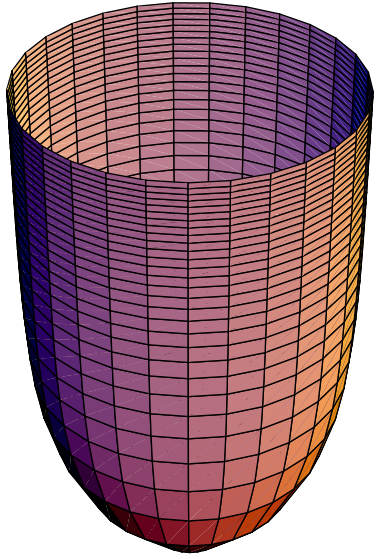,width=7cm}
\vspace{-1.8cm}
\caption{\small \em
The surface $\Xi_{0}$ embedded as a convex surface of revolution 
in $\mathbb{R}^{3}$.
}
\end{center}
\end{figure} 
It follows from
\begin{equation}\label{asymptI}
\lim_{a\rightarrow + \infty}\left(a \sqrt{I(a)}\right)=\sqrt{8 \pi}
\end{equation}
that this surface is asymptotic to a cylinder of radius $\sqrt{8 \pi}$
for large $a$. Moreover,
\[
\lim_{a\rightarrow 0^{+}}\left(a \sqrt{I(a)}\right)=0
\]
implies that it can be completed to a simply-connected surface
by adding the single point at $a=0$. However, this
completion fails to be smooth. One way to see this is to consider the
scalar curvature of the surface; it depends only on $a$, and 
can be easily calculated in terms of elliptic integrals from the formula
(cf.~\cite{Jos})
\[
R(a)=-\frac{1}{2 a I(a)}
\frac{d}{d a}\left( a \frac{d}{da} \log I(a)\right).
\]
We find that this is a positive function on $]0,+\infty[$, monotonically
decreasing, and with limits
\[
\lim_{a\rightarrow 0^{+}}R(a)=+\infty, \qquad
\lim_{a\rightarrow +\infty}R(a)=0;
\]
a plot of $R(a)$ is shown in Figure~5. 
\begin{figure} \label{figcurv}
\begin{center}
\vspace{1cm}
\epsfig{file=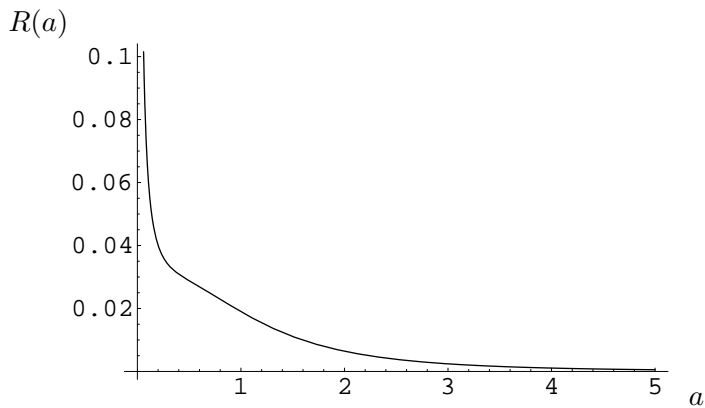,width=8cm}\\
\vspace{-5.5cm}\hspace{-8.75cm}$R(a)$\\
\vspace{4.5cm}\hspace{8.75cm}$a$\\
\vspace{.5cm}
\caption{\small \em
The scalar curvature $R(a)$ on $\Xi_{0}$.
}
\end{center}
\end{figure} 
Thus $\Xi_{0}$ is asymptotically flat for large $a$, which fits with the
asymptotics already mentioned. 
The unboundedness of the curvature as
$a\rightarrow 0$ implies that the one-point completion is not smooth at
the tip $a=0$. A rather surprising fact is that this occurs even though the
profile curves of the surface $\Xi_{0}\hookrightarrow \mathbb{R}^{3}$
approach the symmetry axis at right angles:
\begin{eqnarray} 
\theta_{0}& = & 
{\rm arctan}\; \lim_{a\rightarrow 0^{+}}\frac{d(a u(a))}{dv(a)}\nonumber \\ 
& = & {\rm arctan}\; \lim_{a\rightarrow 0^{+}}
\frac{2\sqrt{I(a)}(2I(a)-aI'(a))}{aI'(a)(4I(a)-aI'(a))}=\frac{\pi}{2}
\label{theta1}
\end{eqnarray}
(here $u$ and $v$ are as defined in the proof of Lemma~\ref{lemembed}). 
Now the limit (\ref{asymptI}) implies that, as $a\rightarrow \infty$, tangents
to the profile curves make an angle of $\theta_{\infty}=0$ with the 
direction of the symmetry axis.
An elementary result on differential geometry of surfaces of revolution in 
$\mathbb{R}^{3}$ then allows us to compute the total curvature 
of $\Xi_{0}$ from (\ref{theta1}) as
\[
\int_{\Xi_{0}}R = 2 \pi (\sin \theta_0 -\sin \theta_{\infty})=2\pi.
\]
This result agrees with what one would obtain for any embedded surface of
revolution in $\mathbb{R}^3$ asymptotic to a cylinder at one end and smooth 
at the other end, by the theorem of Gau\ss--Bonnet.

Any meridian (given by equating $\vartheta$ to
a constant) is a geodesic of the surface $\Xi_{0}$. 
It follows from our proof of
Theorem~\ref{thmincomplete} that the
meridians are incomplete geodesics -- any point on them is at finite
distance from the tip $a=0$. This can also be checked from the
explicit formulas (\ref{metric00}) and (\ref{Iofa}).
It is also easy to show that
the integrand in (\ref{vofa}) never vanishes, and this implies that
none of the parallels (circles of constant $a$) is a geodesic 
(cf.~\cite{Pre}, p.~182). 
More general geodesics on $\Xi_{0}$ are straightforward to find as an 
application of Clairaut's theorem (cf.~\cite{Pre}, pp.~183--185).

\subsubsection{$p=\infty$} \label{secs2linf}

The metric on the totally geodesic submanifold $\Xi_{\infty}$ introduced
in Lemma~\ref{lemtotalgeod} can be written as
\begin{equation}\label{metrpinfty}
\gamma|_{\Xi_{\infty}}=\gamma_{\alpha \bar{\alpha}}(\alpha)|d\alpha|^{2}
\end{equation}
where
\[
\gamma_{\alpha\bar{\alpha}}(\alpha)=2i\int_{\mathbb{C}}
\frac{|w-w^{-1}|^{2}}{\left(|w+\alpha|^2+|w^{-1}+\alpha|^2\right)^{2}}
\frac{dw\wedge d\bar{w}}{|w|^2}.
\]
Notice that the prefactor $\gamma_{\alpha \bar{\alpha}}$ diverges at the
points $\alpha=1$ and $\alpha=-1$, where the degree of the maps 
$z \mapsto \frac{e^{-z}+\alpha}{e^z + \alpha}$ jump to one. Notice also that
$\gamma_{\alpha \bar{\alpha}}$ depends on
both the modulus and the argument of $\alpha$, which makes the computation
of the integral in closed form a more difficult task. However, we can
calculate some geodesics of $\gamma_{\alpha\bar{\alpha}}$ even without
performing the integral.

\begin{lemma} \label{lemgeopinfty}
The intervals $]-\infty, -1[$, $]-1,1[$, $]1,+\infty[$ and 
$i\mathbb{R}$ in the complex plane parametrised by $\alpha$ are all 
geodesics of the metric (\ref{metrpinfty}).
\end{lemma}
\begin{proof}
This follows again from Lemma~\ref{lemisometries}. Invariance of the maps
\[
W(z)=\frac{e^{-z}+\alpha}{e^z + \alpha} \quad \in \Xi_{\infty} 
\]
with respect to the isometry $\tilde{\sigma}_{2}$ defined in
(\ref{sigtil2}) imposes the relation
\[
\alpha=\bar{\alpha};
\]
on $\Xi_{\infty}$, this is the equation for the union of the three
intervals $]-\infty, -1[$, $]-1,1[$ and $]1,+\infty[$, which are therefore
geodesics of $\gamma|_{\Xi_{\infty}}$. Similarly, considering 
$R:w\mapsto w^{-1}$, invariance under the isometry
\[
T_{i\pi}\circ R \circ \tilde{\sigma}_{2}:
\frac{e^{-z}+\alpha}{e^z + \alpha} \mapsto \frac{e^{z}+\bar{\alpha}}
{e^{-z} + \bar{\alpha}}
\]
leads to the constraint
\[
\alpha=-\bar{\alpha},
\]
and this shows that $i\mathbb{R}$ is also a geodesic.
\end{proof}

The proof of Theorem~\ref{thmincomplete} implies that the 
geodesic segment in $\Xi_{\infty}$ corresponding to 
$\alpha \in [\frac{1}{2},1[$, has finite
length with respect to the metric (\ref{metrpinfty}), and the same is true 
for any other piece of the intervals in Lemma~\ref{lemgeopinfty} that 
accumulates at $\alpha=1$ or $-1$.

Analogously to the case of $\Xi_{0}$ above, we can prove that the
scalar curvature of $\Xi_{\infty}$ becomes unbounded in the neighbourhood
of the points where the metric becomes singular:

\begin{lemma} \label{lemunboundRXinf}
The scalar curvature $R$ of $\Xi_{\infty}$ satisfies
\[
\lim_{\alpha \rightarrow \pm 1} R(\alpha) = +\infty.
\]
\end{lemma}
\begin{proof}
We focus on the limit $\alpha \rightarrow +1$ without loss of generality.
Consider the paths $\gamma_{\infty,u}:[\frac{1}{2},1[ \rightarrow
\Xi_{\infty}$, with $u$ in the unit circle of $\mathbb{C}$, given by
\[
\gamma_{\infty,u}(t):w\mapsto 
\frac{(ut-u+1)w +1}
{w(w+ut-u+1)}=\tilde{W}_{u,t}(w).
\]
These paths parametrise radial segments tending to $\alpha=1$. (Notice
that $\gamma_{\infty,1}$ coincides with $\gamma_{\infty}$ defined in
the proof of Theorem~\ref{thmincomplete} (iii)). An analogous argument
to the one in Theorem~\ref{thmincomplete}, and which we shall not 
reproduce here, leads to the following 
estimate for the curvature of each $\gamma_{\infty, u}$:
\begin{eqnarray*}
k_{u}(t) & = &
\frac{\D \frac{\partial}{\partial t} \int_{\mathbb{C}}\Phi_{u}(w,t)\,d^2 w}
{\D 2 \left( \int_{\mathbb{C}} \Phi_{u}(w,t)\,d^2 w \right)^{1/2}} \\
& > & \D C_{u}
\frac{\D \frac{1}{1-t}-\frac{(1-t)^3}{(\varepsilon^2+(1-t)^2)^2}}
{\D \left(
\log \frac{\varepsilon^2 + (1-t)^2}{(1-t)^2}-
\frac{\varepsilon^2}{\varepsilon^2 + (1-t)^2}
\right)^{1/2}}.
\end{eqnarray*}
Here, $\Phi_{u}(w,t)$ is again determined from $\tilde{W}_{u,t}(w)$ by 
(\ref{Phi}) 
and $C_{u}$ denote positive 
constants dependent on $u$ and on a
fixed $\varepsilon \in \,]0,\frac{1}{2}[$. Since the right-hand-side
is strictly positive when $t\rightarrow 1^{-}$, 
the scalar curvature must be positive
in some neighbourhood of $\alpha=1$ by a continuity argument. The
inequality above also implies that the minimal curvature becomes unbounded 
as $t\rightarrow 1^{-}$, and the result follows.
\end{proof}

%-------------------------------------------------------------------------

\subsection{Two-lump scattering} \label{sec2lscat}

Now that we have found some geodesics on ${\cal M}_2$, we can interpret
them in terms of soliton scattering by plotting the energy density 
(\ref{endens}) along them.

%-------------------------------------------------------------------------

\subsubsection{$p=0$} \label{sec2ls0}

We have seen that $\Xi_{0}$ is a surface of revolution. The meridians of
this surface define a one-parameter family of geodesics; one of them is
\[
\Gamma_{0}:=\{ z\mapsto \alpha \sech z : \alpha \in \, ]0,+\infty[\}.
\]
All the other meridians are related to $\Gamma_{0}$ through a fixed target
rotation, under which the energy density (\ref{endens}) does not change,
and therefore describe the same type of process. This process can be
interpreted as a frontal collision of two lumps as we let $\alpha$
decrease from a large value to zero; a plot of the energy densities is
shown in Figure~6. For large $\alpha$, the configuration
can be roughly described as a superposition of two single lumps with
$\SO(2)$ symmetry (thus having antipodal endpoints) which are far apart.
As $\alpha$ decreases, these lumps approach each other (meaning that the 
regions of large ${\cal E}$ come closer together on the cylinder), and at
close distance the approximate $\SO(2)$ symmetry of the energy density
breaks down. At this stage,
energy density peaks form over antipodal points of a circle transverse to 
the axis of the cylinder; these peaks become more and more pronounced, with a
singularity forming in the limit $\alpha \rightarrow 0^{+}$.
As we have seen, this is achieved in finite time,
which is a symptom of the 
incompleteness of the metric. This type of phenomenon is not
surprising for the $\mathbb{CP}^{1}$ model; it has been reproduced in 
numerical studies of scattering lumps on the plane \cite{LeeLC}.

\begin{figure}\label{fig2lump}
\begin{center}
\raisebox{.6cm}{\epsfig{file=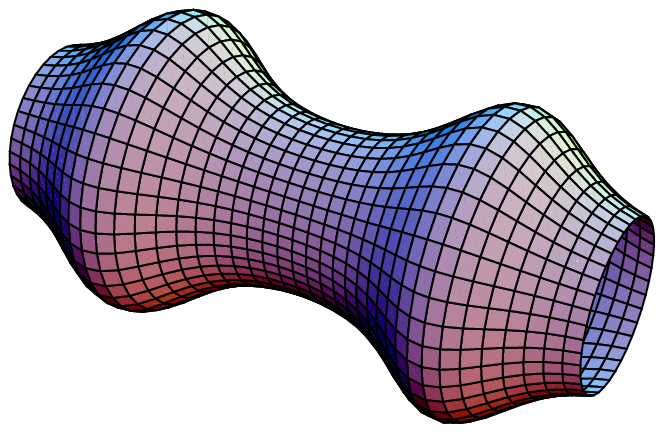,width=3cm}} 
\raisebox{2cm}{$\rightarrow$}
\raisebox{.6cm}{\epsfig{file=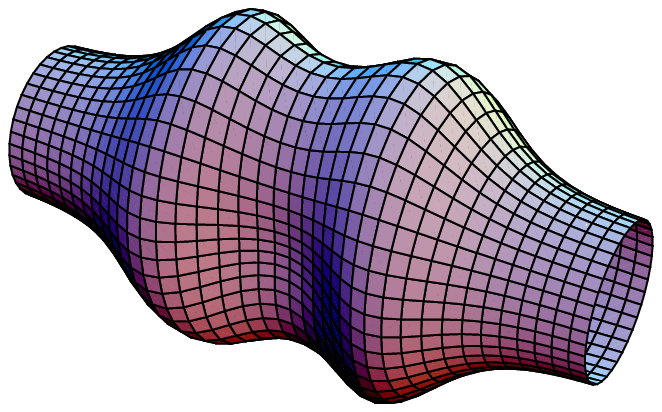,width=3cm}} 
\raisebox{2cm}{$\rightarrow$}
{\epsfig{file=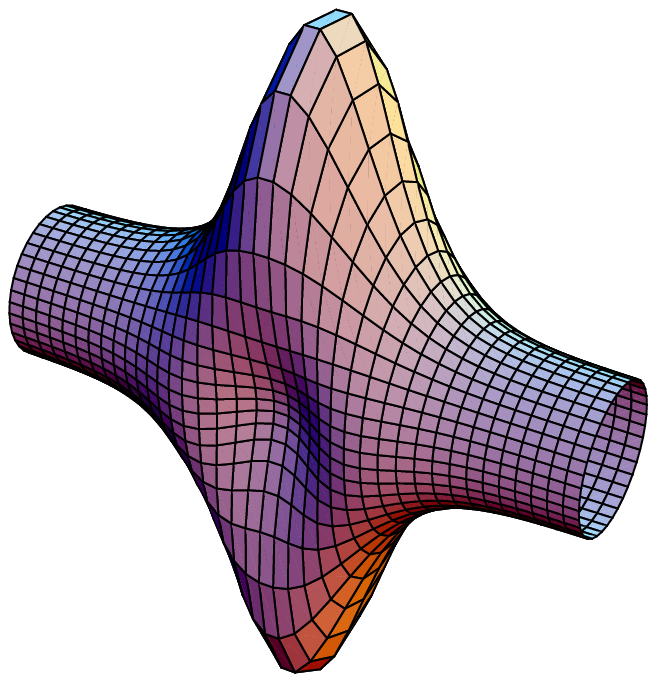,width=3cm}} 
\raisebox{2cm}{$\rightarrow$}
\raisebox{.2cm}{\epsfig{file=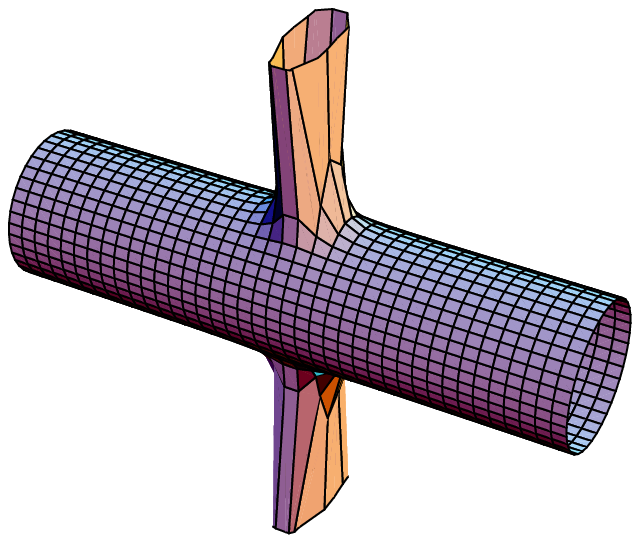,width=3cm}}\\[-.5cm]
(a) \hspace{3cm} (b) \hspace{3cm} (c) \hspace{3cm} (d)
\caption{\small \em
Frontal colision of a symmetric configuration of 2-lumps on
${\cal M}_{2}^{0}$, corresponding to the geodesic $\Gamma_{0}$. 
Energy densities of lumps of the form
$W(z)=\alpha \sech z$ are plotted on top of a cylinder of unit
radius for $|{\rm Re}\,z|\le 4$:
{\em (a)} $\alpha=5$; 
{\em (b)} $\alpha=2$; 
{\em (c)} $\alpha=1$; 
{\em (d)} $\alpha=0.05$.}
\end{center}
\end{figure}

It is a consequence of Clairaut's theorem that the geodesics of $\Xi_{0}$ 
other than the meridians are complete and do not involve singular peaking. 
A simple way to understand them (cf.~\cite{SpeS2,SpeT2})
is to interpret the geodesic flow on 
$\Xi_{0}$ as the dynamics of a particle in $]0,+\infty[$ with
position-dependent mass $I(a)$ and lagrangian 
\[
L=\frac{1}{2}I(a)\dot{a}^{2}+ p_{\vartheta}^{2}\,U_{\rm eff}(a),
\]
where
\begin{equation} \label{Ueff}
U_{\rm eff}(a)=\frac{1}{2 a^2 I(a)}.
\end{equation}
Here, $p_{\vartheta}=a^ 2 I(a) \dot{\vartheta}$ 
is the (conserved) momentum conjugate to the 
cyclic coordinate $\vartheta$, which can be interpreted as a coupling to
the effective potential (\ref{Ueff}). We plot $U_{\rm eff}(a)$ in
Figure~7; it is a monotonically decreasing function and
has a horizontal asymptote at $\frac{1}{16 \pi}$ as 
$a\rightarrow +\infty$, corresponding to the 
limit (\ref{asymptI}). From the plot, it is immediately clear that the
complete geodesics of $\Xi_{0}$, corresponding to taking 
$p_{\vartheta}\ne 0$, describe reflection collisions. In
these processes, the peaking of the 
energy density as the lumps approach each other is reversed at a certain
instant, after which the lump separation grows to infinity. 
For instance, the sequence 
(a)$\rightarrow$(b)$\rightarrow$(c)$\rightarrow$(b)$\rightarrow$(a)
of configurations in Figure~6 represents snapshots of one such reflection.
Processes of this type are accompanied by a rotation of the overall 
phase of the field
configuration and are generic among the motions on the submanifold $\Xi_{0}$.

\begin{figure} \label{figeffpot}
\begin{center}
\vspace{1cm}
\epsfig{file=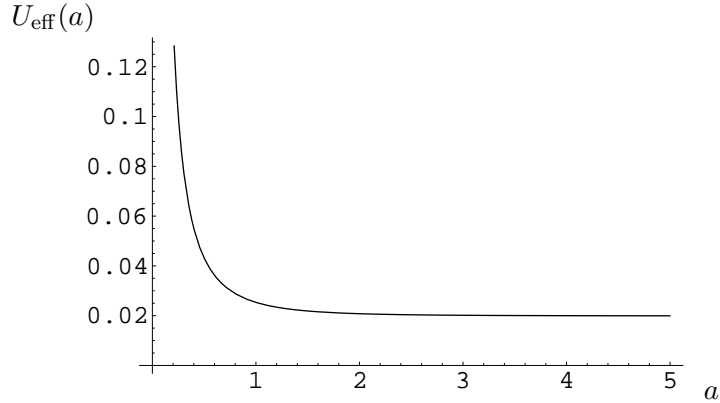,width=8cm}\\
\vspace{-5.5cm}\hspace{-8.75cm}$U_{\rm eff}(a)$\\
\vspace{4.5cm}\hspace{8.75cm}$a$\\
\vspace{.5cm}
\caption{\small \em
The effective potential $U_{\rm eff}(a)$.
}
\end{center}
\end{figure}

%-------------------------------------------------------------------------

\subsubsection{$p=\infty$} \label{sec2lsinf}

The geodesics we have found on ${\cal M}_{2}^{\infty}$ give three 
qualitatively distinct two-lump motions:
\[
\Gamma_{j}:=\left\{ z\mapsto \frac{e^{-z}+\alpha}{e^{z}+\alpha}: 
\alpha \in I_{j} \right\}, 
\qquad j=1,2,3
\]
with
\begin{eqnarray*}
&I_{1}:=]1,+\infty[,&\\
&I_{2}:=i\mathbb{R},&\\
&I_{3}:=]-1,1[.&
\end{eqnarray*}

Energy densities of the process described by $\Gamma_{1}$ are plotted in
Figure~8. We can say it consists of a frontal 
collision of two peaked single lumps of the same shape along a longitudinal 
(straight) line. At collision, the two peaks coalesce and develop a singularity
over the midpoint of their initial positions (local maxima of ${\cal E}$) in 
the limit $\alpha \rightarrow 1^{+}$.

\begin{figure} \label{fig2lumpi}
\begin{center}
\raisebox{0cm}{\epsfig{file=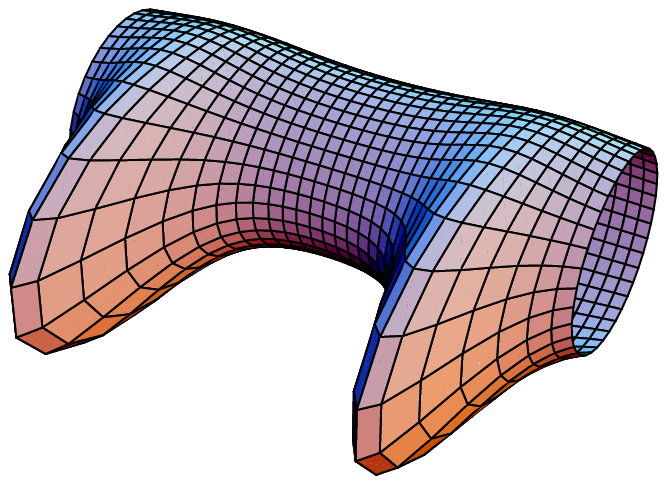,width=3.5cm}} 
\hspace{.4cm}\raisebox{1.6cm}{$\rightarrow$}
\raisebox{-.01cm}{\epsfig{file=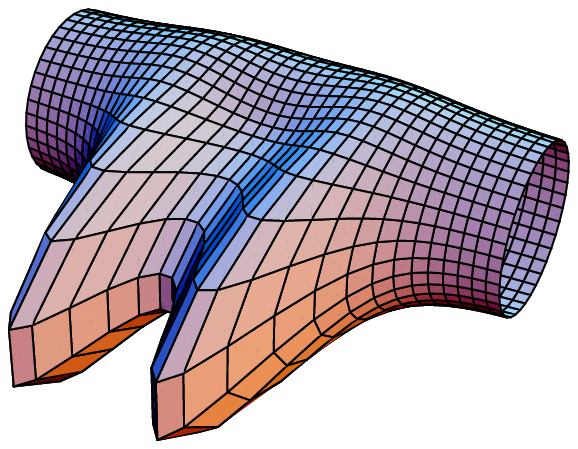,width=3.5cm}} 
\hspace{.4cm}\raisebox{1.6cm}{$\rightarrow$}
\raisebox{-0.5cm}{\epsfig{file=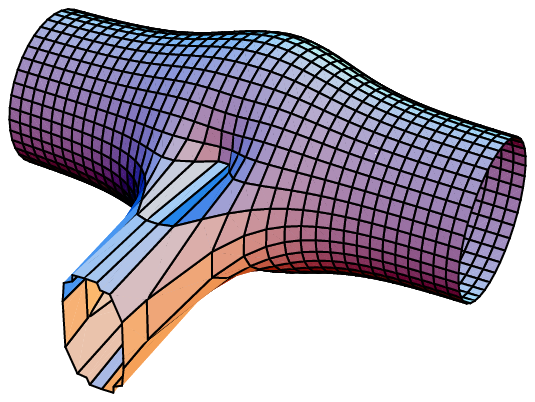,width=3.5cm}} \\[-.7cm]
(a)\hspace{4cm} (b)\hspace{4cm} (c)
\caption{\small \em
Frontal collision of lumps, corresponding to the geodesic $\Gamma_{1}$ of
${\cal M}_{2}^{\infty}$. Energy densities of lumps of the form 
$W(z)=(e^{-z}+\alpha)/(e^{z}+\alpha)$ are plotted on a cylinder
of unit radius for $|{\rm Re}\,z|\le 3$:
{\em (a)} $\alpha=5$; 
{\em (b)} $\alpha=2$; 
{\em (c)} $\alpha=1.1$.}
\end{center}
\end{figure}

The geodesic $\Gamma_{2}$ describes processes related to the decay of the
two-lump $W(z)=e^{-2z}$ (whose energy density exhibits $\SO(2)$ symmetry) to 
peaked configurations that
become singular in the limit $\alpha\rightarrow \mp 1^{\pm}$. 
Strictly speaking, this process alone is not of scattering nature 
because it does
not connect configurations of asymptotically well-separated maxima of
energy density.
Alternatively,
one could interpret the geodesic as a tunneling of single-peaked two-lumps 
through the cylinder, passing through the $\SO(2)$-symmetric configuration.
This is illustrated in Figure~9.

\begin{figure} \label{fig2lumpii}
\begin{center}
\raisebox{0.5cm}{\epsfig{file=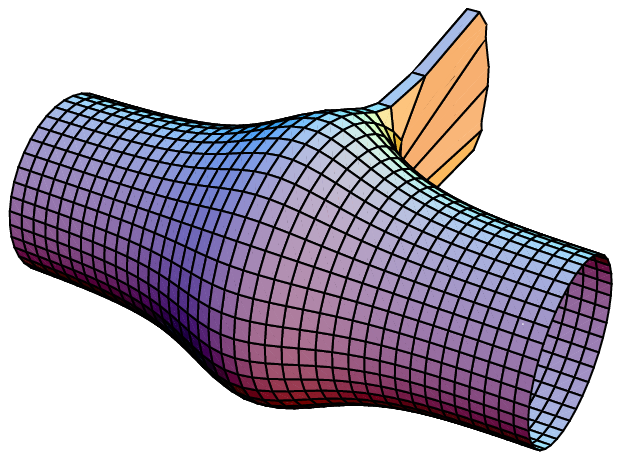,width=2.2cm}} 
\raisebox{1.51cm}{$\rightarrow$}
\raisebox{0.3cm}{\epsfig{file=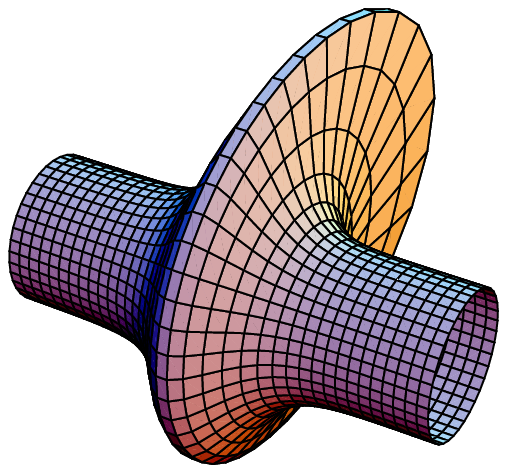,width=2.5cm}} 
\raisebox{1.5cm}{$\rightarrow$}
\raisebox{0cm}{\epsfig{file=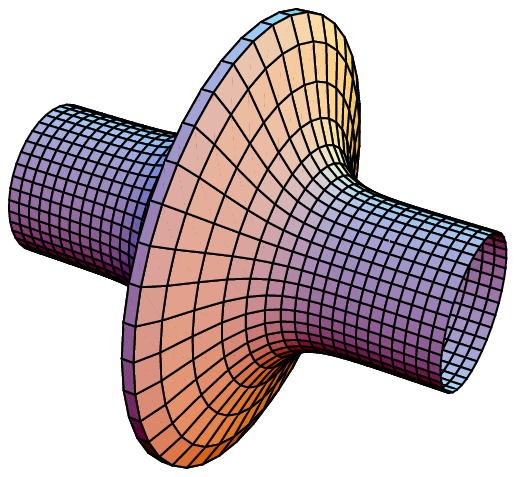,width=2.5cm}} 
\raisebox{1.5cm}{$\rightarrow$}
\raisebox{-0.2cm}{\epsfig{file=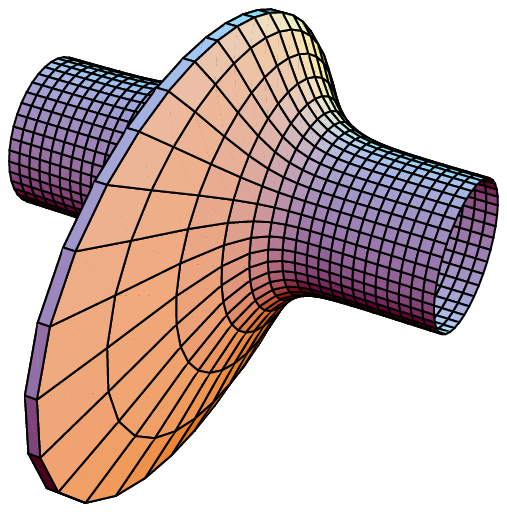,width=2.5cm}}
\raisebox{1.5cm}{$\rightarrow$}
\raisebox{0.3cm}{\epsfig{file=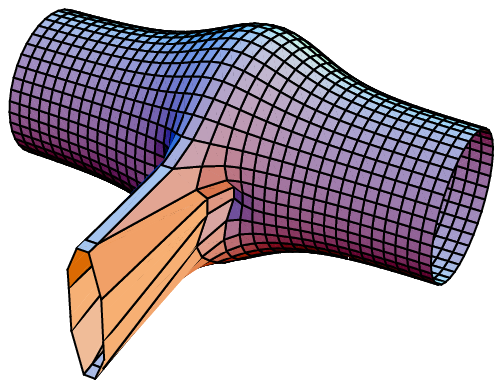,width=2.2cm}} \\[-.3cm]
(a)\hspace{2.6cm}(b)\hspace{2.6cm}(c)\hspace{2.6cm}(d)\hspace{2.6cm}(e)
\caption{\small \em
Tunneling of 2-lumps through the cylinder, 
corresponding to the geodesic $\Gamma_{2}$ of
${\cal M}_{2}^{\infty}$. Energy densities of lumps of the form 
$W(z)=(e^{-z}+\alpha)/(e^{z}+\alpha)$ are plotted on a cylinder
of unit radius for $|{\rm Re}\,z|\le 3$:
{\em (a)} $\alpha=-0.95$; 
{\em (b)} $\alpha=-1/3$; 
{\em (c)} $\alpha=0$;
{\em (d)} $\alpha=1/3$;
{\em (e)} $\alpha=0.95$.
}
\end{center}
\end{figure}

Finally, the geodesic $\Gamma_{3}$ may be interpreted as a scattering process
of two single lumps with the same shape along longitudinal lines positioned 
antipodally on the cylinder; energy densities are plotted in 
Figure~10. If the lumps travel past each other, there is 
again an instant 
for which the energy density of the configuration has $\SO(2)$ symmetry, and
after that each individual lump continues its motion along the longitudinal
line with no significant distortion of shape.

\begin{figure} \label{fig2lumpiii}
\begin{center}
\raisebox{0cm}{\epsfig{file=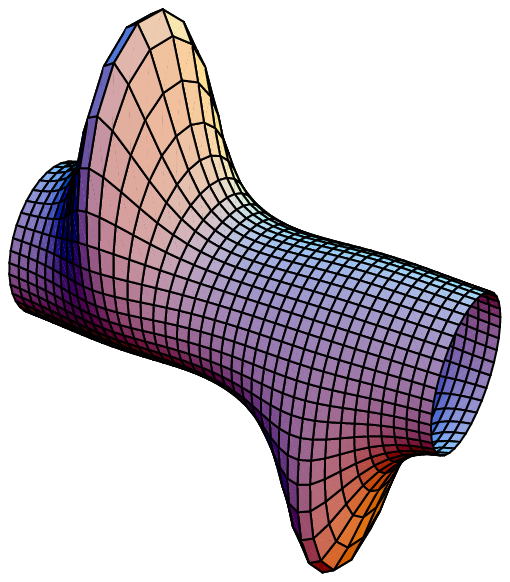,width=3cm}} 
\hspace{.3cm}\raisebox{2.1cm}{$\rightarrow$}
\raisebox{-.2cm}{\epsfig{file=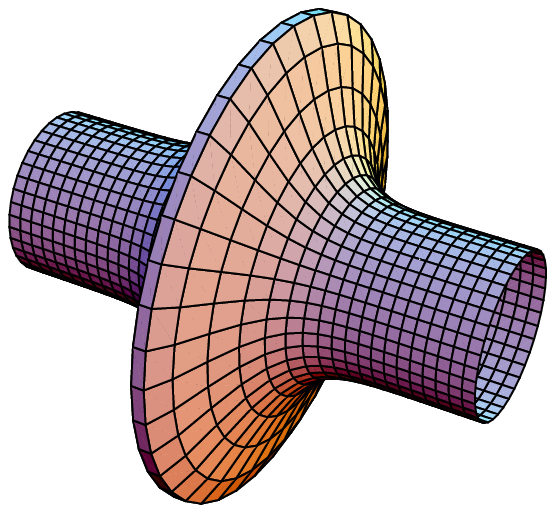,width=4cm}} 
\raisebox{2.1cm}{$\rightarrow$} \hspace{.3cm}
\raisebox{0cm}{\epsfig{file=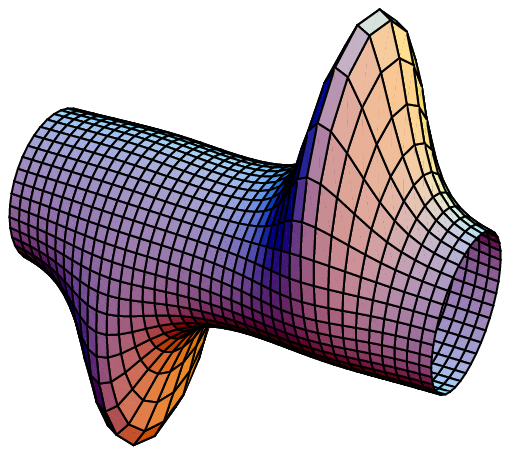,width=3cm}}\\[-.5cm] 
(a)\hspace{4cm}(b)\hspace{4cm}(c)
\caption{\small \em
Scattering of antipodal lumps, corresponding to the geodesic $\Gamma_{3}$ of
${\cal M}_{2}^{\infty}$. Energy densities of lumps of the form 
$W(z)=(e^{-z}+\alpha)/(e^{z}+\alpha)$ are plotted on a cylinder
of unit radius for $|{\rm Re}\,z|\le 3$:
{\em (a)} $\alpha=3i$; 
{\em (b)} $\alpha=0$; 
{\em (c)} $\alpha=-3i$.
}
\end{center}
\end{figure}

%====================================================================

\section{Discussion} \label{secdisc}

In this paper, we have considered the adiabatic approximation
to the dynamics of solitons in the $\mathbb{CP}^1$ 
$\sigma$-model on an infinite
cylinder $\Sigma$. As in previous studies of this model on other surfaces, 
for each degree $n\in\mathbb{Z}$ there is
a smooth, finite-dimensional moduli space ${\cal M}_{n}$ parametrising 
static solutions ($n$-lumps); in our case, this space is
modelled on the space of rational
maps $\mathbb{CP}^{1}\rightarrow\mathbb{CP}^{1}$ and is therefore a complex
manifold. We have found that the approximation
defines a dynamical system by automorphisms of a natural map
$\ell: {\cal M}_{n}\rightarrow \mathbb{CP}^{1}\times\mathbb{CP}^{1}$
specifying the boundary values of the fields. On each fibre, these
automorphisms are defined by the geodesic flow of the $L^{2}$ metric, which
is regular and K\"ahler. By means of this fibration, we avoid making
reference to degenerate metrics as in~\cite{WarSML}. Although
our language could be adapted to deal with lumps on the plane, in that case 
the boundary values of the fields alone are still not enough to specify a 
sufficiently fine fibration of the moduli spaces to render the metrics 
regular.

Lumps of degree one are characterised by a shape function $d$ taking values
in $]0,\pi]$ (the distance of their endpoints), together with a location 
(a point on $\Sigma$ if $d\ne 0$ or a transversal circle if $d=0$) and a 
physically irrelevant global phase. Their adiabatic 
dynamics is trivial: it reduces to uniform motion of their location on the 
cylinder with shape-independent inertial mass. This is similar to
the $\mathbb{CP}^{1}$ $\sigma$-model on the plane, 
where the only possible adiabatic 
motion of one-lumps is also uniform motion along geodesics,
i.e.~straight lines~\cite{WarSML}.

The dynamics of multilumps is more interesting to study. We established
that all the metrics for multilumps on the cylinder are incomplete. Again,
this parallels an analogous result for lumps on the plane, as put forward
by Sadun and Speight in~\cite{SadSpe}. Incompleteness of the metric
translates into the possibility of lump collapse in finite time in the
adiabatic approximation.
Using standard symmetry considerations, we have found
totally geodesic submanifolds for the metrics on two types of fibres 
${\cal M}_{2}^{(p,q)}=\ell^{-1}(\{p,q\})$, namely for $p=q$ and $p,q$
antipodal, and some geodesics on them. We have also found explicit formulae
for the metric on one two-dimensional totally geodesic submanifold of
${\cal M}_{2}^{0} \cong {\cal M}_{2}^{(p,p)}$, 
which involves elliptic integrals. 
This metric is incomplete, and the corresponding lump collapse
can be plotted with no difficulty (Figure~6).
Similarly, some of the geodesics we found for $p$ and $q$ antipodal exhibit
lump collapse (Figures 8 and 9).
It is still an unsettled question how to interpret finite-time collapse 
(which is understood as a feature of the adiabatic approximation) at the 
full field theory level. As the metric becomes singular, one may expect the 
approximation to break down; on the other hand, numerical simulations of
the field theory seem to support the claim that collapse in finite time 
should also occur in the full dynamics. Another question is whether the
dynamics is well defined beyond collapse. A rather striking feature of
the geodesics describing collapse that we found is that they all have natural
prolongations on the relevant moduli spaces; in particular, for the 
scattering processes we described on 
$\Xi_{\infty}$, the whole real $\alpha$-axis can be interpreted as a
process of double scattering at $90^{\rm o}$ of two one-lumps approaching 
first along a generatrix of the cylinder, then travelling along a 
transversal circle, 
and finally separating along the opposite generatrix, which is very natural 
to expect from our intuition on (second-order) soliton dynamics in two 
dimensions.

The scattering processes corresponding to the geodesics that we found 
explicitly turn out to be rather unique when compared to previous 
results on other surfaces, a fact that is due to the different topology 
of the cylinder. It should be expected that more generic geodesics will 
give rise to more familiar processes, in particular the frontal scattering at 
$90^{\rm o}$. In fact, we have found 
curves on the submanifold $\tilde{\Xi}_{0}$ which are close to geodesics 
(in the sense that the Christoffel symbols related to transverse motion
are small in some region) and that describe processes of this type.

There is some belief that lump configurations at collapse 
are supressed at the quantum level. Following Gibbons and 
Manton~\cite{GibManCQM}, the quantum-mechanical version
of the adiabatic dynamics should be based on a Schr\"odinger equation on 
each ${\cal M}_{n}^{(p,q)}$ using the covariant laplacian of the $L^2$
metric, but as a correction one expects an
effective potential term given by the scalar curvature of the moduli
space~\cite{MosShi}. 
Accordingly, wavefunctions should be given zero boundary values
whenever the scalar curvature diverges.
In our examples, we found that the scalar curvatures of $\Xi_{0}$ and
$\Xi_{\infty}$ blow up as the boundary of the moduli space is approached 
upon collapse. 
In \cite{SpeL2}, Speight also found a divergence of the
scalar curvature of the moduli space of one-lumps on $S^2$ preventing
collapse, but our results directly refer to the interacting case and 
therefore give more substantial support to the hope that the degree of 
lumps should be conserved in the quantum field theory.

%====================================================================

\vspace{1cm}
\noindent
{\large \bf{Acknowledgements}}
\vspace{.5cm}

\noindent
I would like to thank Klaus Kirsten, Nick Manton, Avijit Mukherjee, 
Oliver Schn\"urer and especially Martin Speight for discussions.
This work was supported by the Max-Planck-Gesellschaft, Germany.

%==============================================================

\begin{appendix}
\section*{Appendix A: Proof of Lemma \ref{lemisometries}} 
\label{apisometries}

Totally geodesic submanifolds $N \subset M$ can be characterised by the
property that (the continuation of) any geodesic of the ambient metric
starting tangent to $N$ will never leave $N$. Suppose, for a contradiction,
that there is a geodesic $r:\,]-\varepsilon, \varepsilon[ \rightarrow M$ of
$M$ such that
\begin{equation} \label{risinF}
r(t) \in F \quad \forall t \in \,]-\varepsilon, 0]
\end{equation}
and
\begin{equation} \label{risnotinF}
r(t) \not\in F \quad \forall t \in \,]0,\varepsilon[.
\end{equation}
It follows from (\ref{risinF}) that $r'(0) \in T_{r(0)}F \subset
T_{r(0)}M$, and as such it can be regarded as an equivalence class of
paths through $r(0)$ containing paths that lie entirely on $F$. These
paths are fixed by $S$, so $f_{\ast r(0)}r'(0)=r'(0)$ for all 
$f \in S$. Now (\ref{risnotinF}) implies that for any $\tilde{t}\in
\,]0,\varepsilon[$  there is at least one
element $\tilde{f} \in S$ such that $\tilde{f}(r(\tilde{t}))
\ne r(\tilde{t})$. Since $\tilde{f} \in \Iso(M)$, 
$\tilde{f}\circ r$ is also a solution to the equation of the geodesics
for $(M,g)$; it satisfies the Cauchy data
\[
\left\{
\begin{array}{l}
(\tilde{f}\circ r)(0)=r(0) \\
(\tilde{f}\circ r)'(0)=r'(0)
\end{array}
\right.
\]
and is distinct from $r$. This contradicts the Picard--Lindel\"of theorem
ensuring uniqueness of solutions of ODEs. Hence $F$ must be a totally
geodesic submanifold. \hfill $\Box$

%=======================================================================

\section*{Appendix B: Proof of Lemma \ref{lemintegr}}
 \label{apintegr}

The fact that $f$ in (\ref{functionf}) is analytic for 
positive real values of $t$ follows
from the (complex) analyticity of the integrand as a function of $t$,
the integrability of the integrand as a function of $k$ and the Leibniz 
rule. 
To compute $f(t)$ in closed form, we use an argument based on the
idea that $f$ can be extended by analytic continuation to a neighbourhood 
of the set
\[
\{ t \in \mathbb{C^{*}}: {\rm Re}\, t\ge 0,\, {\rm Im}\,t \ge 0\}.
\]

We start by rewriting
\begin{eqnarray*}
f(t) & = &
\int_{0}^{1}\left( \frac{k-t}{k^{2}-t^{2}} + 
\frac{1-tk}{1-t^{2}k^{2}} \right) K(k)dk \\
 & = &
\int_{0}^{1}\left( \frac{k}{k^{2}-t^{2}}+\frac{1}{1-t^{2}k^{2}}\right)
K(k)dk + \frac{1}{t}\int_{0}^{1}
\left( \frac{k}{k^{2}-t^{-2}}+\frac{1}{1-t^{-2}k^{2}}\right)
K(k)dk;
\end{eqnarray*}
here, the last two integrals must be interpreted as Cauchy
principal values, which are easily seen to exist. To evaluate their
sum, we first Wick-rotate $t$ to $it$, which leads to
\[
\int_{0}^{1}\left( \frac{k}{k^{2}+t^{2}}+\frac{1}{1+t^{2}k^{2}} \right)
K(k)dk -\frac{i}{t}\int_{0}^{1}\left( \frac{k}{k^{2}+t^{-2}}+
\frac{1}{1+t^{-2}k^{2}}\right)K(k)dk.
\]
Each of the two terms above can be evaluated in closed form using the
result (cf. formula I.(5) in \cite{Gla})
\[
\int_{0}^{1}\left( \frac{k}{k^{2}+z^{2}} + \frac{1}{1+z^{2}k^{2}}\right)
K(k) dk=\frac{1}{\sqrt{1+z^{2}}}K\left(\frac{1}{\sqrt{1+z^{2}}}\right).
\]
This yields
\begin{equation} \label{stillWickrot}
\frac{1}{\sqrt{1+t^{2}}}K\left(\frac{1}{\sqrt{1+t^{2}}}\right)
-\frac{i}{t}\frac{1}{\sqrt{1+t^{2}}}K\left(\frac{1}{\sqrt{1+t^{2}}}\right).
\end{equation}

We have to now undo the Wick rotation in the expression above 
to obtain the values of $f$ we are interested in. This must be done
carefully, since the analytic continuation of $K$ branches at the 
singular point $k=1$ and the square root branches at the origin.
Recall that $K(k)$ can be represented as a hypergeometric series 
for $0<k<1$ (cf.~(900.00) in \cite{ByrFri}):
\begin{equation} \label{seriesK}
K(k)=\frac{\pi}{2} 
\left.\right._{2}\! F_{1}\left( \frac{1}{2},\frac{1}{2};1;k^{2}\right)=
\frac{\pi}{2}\sum_{j=0}^{\infty}
\left(\frac{(2j-1)!!}{2^{j}j!}\right)^{2}k^{2}.
\end{equation}
(Here $(2n-1)!!:=(2n-1)(2n-3)\cdots 1$ and $(-1)!!:=1$.)
So the properties of the analytic continuation of $K$ can be deduced 
from those of Gau\ss's $_{2}F_{1}$.
Following common practice, we
introduce a branch cut on the real axis from $1$ to $+\infty$. On
$\mathbb{C}-[1,+\infty]$, $K$ is single-valued, and it commutes with
complex conjugation,
\begin{equation} \label{Kcommcc}
K(\overline{k})=\overline{K(k)},
\end{equation}
because the coefficients of the series in (\ref{seriesK}) are real.
Across the branch cut, there is a nontrivial monodromy that accounts 
for a discontinuity
\begin{equation} \label{discK}
K(k)=\frac{1}{k}\left(
K\left(\frac{1}{k}\right)\pm 
i K\left( \frac{\sqrt{k^{2}-1}}{k}\right) \right), 
\qquad k\in \,]1,+\infty[,
\end{equation}
where the top/bottom signs correspond to the limits obtained when
$k$ approaches the cut from above/below. This result can
be obtained by relating Kummer's solutions of hypergeometric differential 
equations (cf.~\cite{Iwa}, Chap.~2). By a similar argument
(and (\ref{Kcommcc})), one can show that
\begin{equation} \label{Kforimag}
K(ik)=\frac{1}{\sqrt{1+k^{2}}}K\left(\frac{1}{\sqrt{1+k^{2}}}\right), 
\qquad k\in \mathbb{R}.
\end{equation}

We now want to Wick-rotate $t$ back to $-it$ in (\ref{stillWickrot}); 
one way to keep track of the branching of the functions involved
is to substitute the final $t$ by $e^{\frac{i\epsilon}{2}}t$, 
with $\epsilon$ 
small, real and positive, evaluate in terms of continuous quantities
and let $\epsilon \rightarrow 0^{+}$ at the end. It is convenient to 
consider the cases $0<t<1$ and $t>1$ separately. In the first case,
we find that the argument of the first $K$ in (\ref{stillWickrot})
after substution,
$(1-e^{i\epsilon}t^{2})^{-1/2}$, has positive imaginary part, and
according to (\ref{discK}) above we should then evaluate
\[
K\left(\frac{1}{\sqrt{1-e^{i\epsilon}t^{2}}}\right) 
\rightarrow \sqrt{1-t^{2}}
\left( K\left( \sqrt{1-t^{2}}\right)+iK(t) \right), \qquad 0<t<1.
\]
On the other hand, the second term in (\ref{stillWickrot}) is free
from branching, and we can evaluate using (\ref{Kcommcc}) and 
(\ref{Kforimag})
\[
K\left( -\frac{i}{\sqrt{t^{-2}-1}}\right)=t\sqrt{t^{-2}-1}K(t),
\qquad 0<t<1.
\]
In the case $t>1$, the argument of the first $K$ 
in (\ref{stillWickrot}) does not lie on the branch cut after rotation,
but the square root itself branches on the negative real half-axis,
meaning that we should take
\[
\frac{1}{\sqrt{1-e^{i\epsilon}t^{2}}} \rightarrow
+\frac{i}{t \sqrt{1-t^{-2}}}, \qquad{t>1}
\]
(where of course $\sqrt{\cdot}$ always denotes the principal branch
of the square root);
we then find using (\ref{Kforimag})
\[
K\left( \frac{i}{\sqrt{t^{2}-1}}\right)=
\sqrt{1-t^{-2}}K
\left(\frac{1}{t}\right), \qquad t>1.
\]
However, this time the second $K$ does branch; since 
$(1-e^{-i\epsilon}t^{-2})^{-1/2}$ has negative imaginary part,
we should take the lower sign in (\ref{discK}) and find
\[
K\left(\frac{1}{\sqrt{1-e^{-i\epsilon}t^{-2}}} \right)
\rightarrow \sqrt{1-t^{-2}}
\left( K\left(\sqrt{1-t^{-2}}\right)-iK\left(\frac{1}{t}\right)\right).
\]
Adding the two terms in each of the two cases $0<t<1$
and $t>1$, we finally obtain the result~(\ref{closedf}). \hfill $\Box$

\end{appendix}

\begin{small}

\bibliographystyle{numsty}
\bibliography{biblio}

%
% This file was generated by bibtex using the style 'hpall.bst'.
%

%
% This is the preamble.
%


%
% These definitions are used by this particular style.
%

\newcommand{\href}[1]{}%
\newcommand{\dhref}[1]{}%
\newenvironment{hpabstract}{%
  \renewcommand{\baselinestretch}{0.2}
  \begin{footnotesize}%
}{\end{footnotesize}}%
\newcommand{\hpeprint}[1]{%
  \href{http://arXiv.org/abs/#1}{\texttt{#1}}}%
\newcommand{\hpspires}[1]{%
  \dhref{http://www.slac.stanford.edu/spires/find/hep/www?#1}{\ (spires)}}%

\begin{thebibliography}{999}
\bibitem{ManRk}
\textsc{N.S. Manton}: A remark on the scattering of BPS monopoles.
\newblock \textsl{Phys. Lett. B} \textbf{110} (1982) 54--56.
%%CITATION = NONE;%%

\bibitem{SamVS}
\textsc{T.M. Samols}: Vortex scattering.
\newblock \textsl{Commun. Math. Phys.} \textbf{145} (1992) 149--180.
%%CITATION = NONE;%%

\bibitem{StuAH}
\textsc{D.~Stuart}: Dynamics of Abelian Higgs vortices in the near Bogomolny
  regime.
\newblock \textsl{Commun. Math. Phys.} \textbf{159} (1994) 51--91.
%%CITATION = NONE;%%

\bibitem{AtiHitMM}
\textsc{M.F. Atiyah {\upshape and} N.J. Hitchin}: The Geometry and Dynamics of
  Magnetic Monopoles.
\newblock Princeton University Press, 1988.
%%CITATION = NONE;%%

\bibitem{StuYMH}
\textsc{D.M.A. Stuart}: The geodesic approximation for the Yang--Mills--Higgs
  equations.
\newblock \textsl{Commun. Math. Phys.} \textbf{166} (1994) 149--190.
%%CITATION = NONE;%%

\bibitem{Nam}
\textsc{M.~Namba}: Families of Meromorphic Functions on Compact Riemann
  Surfaces (Lecture Notes in Mathematics 767).
\newblock Springer-Verlag, 1979.
%%CITATION = NONE;%%

\bibitem{GibManCQM}
\textsc{G.W. Gibbons {\upshape and} N.S. Manton}: Classical and quantum
  dynamics of BPS monopoles.
\newblock \textsl{Nucl. Phys. B} \textbf{274} (1986) 183--224.
%%CITATION = NONE;%%

\bibitem{SchQSM}
\textsc{B.J. Schroers}: Quantum scattering of BPS monopoles at low energy.
\newblock \textsl{Nucl. Phys. B} \textbf{367} (1991) 177--214.
%%CITATION = NONE;%%

\bibitem{RomQCS}
\textsc{N.M. Rom\~ao}: Quantum Chern--Simons vortices on a sphere.
\newblock \textsl{J. Math. Phys.} \textbf{42} (2001) 3445--3469,
  \hpeprint{hep-th/0010277}.
%%CITATION = NONE;%%

\bibitem{WarSML}
\textsc{R.S. Ward}: Slowly-moving lumps in the $\mathbb{CP}^{1}$ model in (2+1)
  dimensions.
\newblock \textsl{Phys. Lett. B} \textbf{158} (1985) 424--428.
%%CITATION = NONE;%%

\bibitem{SpeS2}
\textsc{J.M. Speight}: Low-energy dynamics of a $\mathbb{CP}^{1}$ lump on the
  sphere.
\newblock \textsl{J. Math. Phys.} \textbf{36} (1995) 796--813,
  \hpeprint{hep-th/9712089}.
%%CITATION = NONE;%%

\bibitem{SpeT2}
\textsc{J.M. Speight}: Lump dynamics in the $\mathbb{CP}^{1}$ model on the
  torus.
\newblock \textsl{Commun. Math. Phys.} \textbf{194} (1998) 513--539,
  \hpeprint{hep-th/9707101}.
%%CITATION = NONE;%%

\bibitem{SpeStr}
\textsc{J.M. Speight {\upshape and} I.A.B. Strachan}: Gravity thaws the frozen
  moduli of the $\mathbb{CP}^{1}$ lump.
\newblock \textsl{Phys. Lett. B} \textbf{457} (1999) 17--22,
  \hpeprint{hep-th/9903264}.
%%CITATION = NONE;%%

\bibitem{SadSpe}
\textsc{L.A. Sadun {\upshape and} J.M. Speight}: Geodesic incompleteness in the
  $\mathbb{CP}^{1}$ model on a compact Riemann surface.
\newblock \textsl{Lett. Math. Phys.} \textbf{43} (1998) 329--334,
  \hpeprint{hep-th/9707169}.
%%CITATION = NONE;%%

\bibitem{DinZakK}
\textsc{A.M. Din {\upshape and} W.J. Zakrzewski}: Skyrmion dynamics in 2+1
  dimensions.
\newblock \textsl{Nucl. Phys. B} \textbf{259} (1985) 667--676.
%%CITATION = NONE;%%

\bibitem{RubSusy}
\textsc{P.J. Ruback}: Sigma model solitons and their moduli space metrics.
\newblock \textsl{Commun. Math. Phys.} \textbf{116} (1988) 645--658.
%%CITATION = NONE;%%

\bibitem{SpeL2}
\textsc{J.M. Speight}: The $L^{2}$ geometry of spaces of harmonic maps $S^{2}
  \rightarrow S^{2}$ and $\mathbb{RP}^{2} \rightarrow \mathbb{RP}^{2}$,.
\newblock \textsl{J. Geom. Phys.} \textbf{47} (2003) 343--368,
  \hpeprint{math.DG/0102038}.
%%CITATION = NONE;%%

\bibitem{HasSpe}
\textsc{M.~Haskins {\upshape and} J.M. Speight}: The geodesic approximation for
  lump dynamics and coercivity of the Hessian for harmonic maps.
\newblock \textsl{J. Math. Phys.} \textbf{44} (2003) 3470--3494,
  \hpeprint{hep-th/0301148}.
%%CITATION = NONE;%%

\bibitem{BelPol}
\textsc{A.A. Belavin {\upshape and} A.M. Polyakov}: Metastable states of
  two-dimensional isotropic ferromagnets.
\newblock \textsl{JETP Lett.} \textbf{22} (1975) 245--247.
%%CITATION = NONE;%%

\bibitem{Lic}
\textsc{A.~Lichnerowicz}: Applications harmoniques et vari\'et\'es
  k\"ahleriennes.
\newblock \textsl{Symp. Math. Bologna} \textbf{3} (1970) 341--402.
%%CITATION = NONE;%%

\bibitem{Rem}
\textsc{R.~Remmert}: Funktionentheorie 2.
\newblock Springer-Verlag, 2nd edition, 1995.
%%CITATION = NONE;%%

\bibitem{LanRFA}
\textsc{S.~Lang}: Real and Functional Analysis.
\newblock Springer-Verlag, 3rd edition, 1993.
%%CITATION = NONE;%%

\bibitem{CarRG}
\textsc{M.P. do~Carmo}: Riemannian Geometry.
\newblock Birkh\"auser, 1992.
%%CITATION = NONE;%%

\bibitem{SutBM}
\textsc{P.M. Sutcliffe}: BPS monopoles.
\newblock \textsl{Int. J. Mod. Phys. A} \textbf{12} (1997) 4663--4705,
  \hpeprint{hep-th/9707009}.
%%CITATION = NONE;%%

\bibitem{ByrFri}
\textsc{P.F. Byrd {\upshape and} M.D. Friedman}: Handbook of Elliptic Integrals
  for Engineers and Physicists.
\newblock Springer-Verlag, 1954.
%%CITATION = NONE;%%

\bibitem{Tri}
\textsc{F.~Tricomi}: Elliptische Funktionen.
\newblock Akademische Verlagsgesellschaft Geest \& Portig, Leipzig, 1948.
%%CITATION = NONE;%%

\bibitem{Jos}
\textsc{J.~Jost}: Differentialgeometrie und Minimalfl\"achen.
\newblock Springer-Verlag, 1994.
%%CITATION = NONE;%%

\bibitem{Pre}
\textsc{A.~Pressley}: Elementary differential geometry.
\newblock Springer-Verlag, 2001.
%%CITATION = NONE;%%

\bibitem{LeeLC}
\textsc{R.A. Leese}: Low-energy scattering of solitions in the
  $\mathbb{CP}^{1}$ model.
\newblock \textsl{Nucl. Phys. B} \textbf{344} (1990) 33--72.
%%CITATION = NONE;%%

\bibitem{MosShi}
\textsc{I.G. Moss {\upshape and} N.~Shiiki}: Quantum mechanics on moduli
  spaces.
\newblock \textsl{Nucl. Phys. B} \textbf{565} (2000) 345--362,
  \hpeprint{hep-th/9904023}.
%%CITATION = NONE;%%

\bibitem{Gla}
\textsc{M.L. Glasser}: Definite integrals of the complete elliptic integral
  $K$.
\newblock \textsl{J. Res. Nat. Bur. Standards Sect. B} \textbf{80B} (1976)
  313--323.
%%CITATION = NONE;%%

\bibitem{Iwa}
\textsc{K.~Iwasaki, H.~Kimura, S.~Shimomura {\upshape and} Y.~Masaaki}: From
  Gauss to Painlev\'e.
\newblock Vieweg, 1991.
%%CITATION = NONE;%%

\end{thebibliography}

\end{small}

\end{document}